\documentclass{aa}  
\usepackage{natbib}
\usepackage{multirow} 
\usepackage{enumitem}
\usepackage{lscape}
\usepackage{hyperref}
\bibpunct{(}{)}{;}{a}{}{,} 

\usepackage{graphicx}
\usepackage{txfonts}

\begin{document}

    \title{Interferometric view into RT Pav's long secondary period}
    \subtitle{Binary versus oscillatory convective modes \thanks{Based on observations collected at the European Southern Observatory (ESO) under programs: 109.23K5.001, 109.23K5.004, 109.23K5.0067, 109.23K5.008, 109.23K5.009.}}

   \author{B. Courtney-Barrer\inst{1,2}, X. Haubois\inst{1}, P. Wood\inst{2}, D. Dionese\inst{4,5}, L. Decin\inst{3}, C. Paladini\inst{1}, I. El Mellah\inst{6,7}, D. Defr\`ere\inst{3}, M. Ireland\inst{2}
          }

   \institute{European Southern Observatory, Alonso de Cordova 3107 Vitacura, 19001, Santiago, Chile \\ \email{benjamin.courtney-barrer@anu.edu.au}
         \and
             Research School of Astronomy and Astrophysics, Australian National University, Canberra 2611, Australia
          \and
              Institute of Astronomy, KU Leuven, Celestijnenlaan 200D, 3001 Leuven, Belgium
          \and 
            Institut d’Astronomie et d’Astrophysique, Université Libre de Bruxelles
            (ULB), CP 226, 1050 Brussels, Belgium
          \and 
            Astronomical Observatory, University of Warsaw, Al. Ujazdowskie 4,
            00-478 Warsaw, Poland
          \and
            Departamento de Física, Universidad de Santiago de Chile, Av. Victor Jara 3659, Santiago, Chile
          \and
            Center for Interdisciplinary Research in Astrophysics and Space Exploration (CIRAS), USACH, Santiago, Chile
             }

   \date{Received February 4, 2025; accepted November 22, 2025}

    \abstract
    {Long secondary periods (LSPs) occur in roughly one third of evolved stars, yet their origin remains uncertain. Two leading hypotheses are oscillatory convective modes and a binary companion enshrouded in dust.} 
    {We investigate the LSP in the red giant RT~Pav using multiwavelength interferometry to test these competing hypotheses.}
    {Observations of RT~Pav were obtained with the VLTI instruments PIONIER, GRAVITY, and MATISSE spanning 1.5–5.0\,$\mu$m, near the expected phase of maximum projected separation under a binary hypothesis. These data were complemented by photometric data and Gaia~DR3 astrometry to constrain companion mass, orbital geometry, and photometric amplitude. Monte Carlo simulations evaluated expected interferometric signatures under both scenarios. Parametric models, including uniform-disk, limb-darkened, uniform-ellipse, binary, and oscillatory convective dipole representations, were fitted to squared-visibility and closure-phase data, informing image reconstructions.}
    {Gaia constrains any potential companion to a mass whose Roche-lobe volume is smaller than the minimum extent required by the observed photometric modulation, implying that any obscuring or scattering region capable of producing the observed variability would lie beyond the gravitationally bound zone of such a companion.  Binary models often return the lowest $\chi^2_\nu$, yet fitted positions are not consistent across wavelength, closure phases do not increase with wavelength as a dusty companion would predict, and we only find significant ($>3\sigma$) detections occurring in two of the four tested instrumental wavebands, which is inconsistent with a coherent companion signal. Furthermore simulations and theoretical estimates indicate that a companion with a $\sim$1\,\% flux ratio, at LSP-consistent separations should be consistently detectable (near or above our $3\sigma$ limits) for standard O-rich asymptotic giant branch (AGB) dust via scattering and/or thermal emission, which is not found. Conversely, an oscillatory convective dipole with a $\sim$200\,K temperature contrast reproduces the H band morphology and the visible light-curve amplitude without violating Gaia or photometric constraints. Finally, significant short wavelength companion signals are completely removed when fitting the residuals of the best fit dipole model.}
    {Our interferometric snapshot of RT Pav, acquired near the phase of maximum projected separation under the binary hypothesis, supports oscillatory convective modes as the most physically consistent explanation for its LSP. A logical next step will be time-resolved spectro-interferometric monitoring across the LSP cycle.}


    \authorrunning{name(s) of author(s)}
   \maketitle

\section{Introduction}
One third of evolved stars exhibit long period oscillations in their light curves, known as long secondary periods (LSPs), and there is still no consensus on what causes this \citep{takayama_2020_LSP_thermal_dipole,pawlak_2021_connection_LSP_RGB_evolution, soszyski_2021_binary_origin_LSP}. These stars occupy a certain region in the period-luminosity (PL) plane known as sequence D \citep{1999_wood_LPV_PL_classification}. The physical origin causing all other sequences in the PL plane of long period variables (LPVs) are known \citep{trabucchi_2019_PL_puls_modes,nicholls_2009}, making the physical origin of sequence D one of the largest gaps in our understanding of stellar variability. 
Since its discovery by \cite{1999_wood_LPV_PL_classification}, the LSP phenomenon has been well characterized for more than 20 years, during which radial velocity (RV) and photometry measurements have been used to propose and constrain various hypotheses \citep{woods_2004_origin_of_LSP,nicholls_2009,soszyski_2021_binary_origin_LSP}. These studies have also highlighted the unique properties of sequence D stars, distinguishing them from the other PL sequence groups \citep{Nicholls_2010_ellipsoidal_variability_sED}. Currently, the two most favored hypotheses based on evidence are: 
\vspace{-2mm}
\begin{itemize}
    \setlength\itemsep{0em}
    \item Oscillatory convective mode hypothesis: LSPs are caused by a dipole oscillatory convective mode.
    \item Binary hypothesis: LSPs are caused by a lower main-sequence or substellar companion enshrouded in a dust cloud. 
\end{itemize}
\vspace{-2mm}
The primary evidence supporting the binary hypothesis was originally the typically steeper decline than rise of the light curve due to a hypothesized cometary structure of the dust cloud, an amplitude distribution consistent with eclipsing dust clouds, and an independence between the observed LSPs and the other pulsation periods. 
Good fits between observed RV curves and binary models \citep{nicholls_2009, woods_2004_origin_of_LSP,1999_wood_LPV_PL_classification} further supported the binary hypothesis, but the eccentricity and angle of periastron distribution muddied this interpretation \citep{nicholls_2009}. In particular, \citet{nicholls_2009} found that if velocity variations in LSP variables are due to binary motion, the measured distribution of the fitted angle of periastron in a large sample of stars has a probability of less than 0.002\% that it comes from randomly aligned binary orbits. More recently, a significant number of secondary dips in LSP light curves, observed in the L and M bands, were interpreted as secondary eclipses where a dusty companion moves behind the primary star \citep{soszyski_2021_binary_origin_LSP}. This led to the strong claim that binarity is the origin of LSPs. Radial velocity measurements of LSP stars initially suggested eccentric orbits with semi-major axes between 1–10 AU and companions with masses consistent with brown dwarfs or lower main-sequence stars \citep{woods_2004_origin_of_LSP, nicholls_2009}. However, recent studies suggest that 85\% of LSP stars likely have circular orbits, with only 15\% exhibiting eccentric orbits \citep{soszyski_2021_binary_origin_LSP}. This finding has been reconciled with earlier results by \citet{Soszynski_2014_rositter}, which showed that non-sinusoidal RV curves in circular orbits could be explained by the Rossiter-McLaughlin effect. For Betelgeuse, \citet{goldberg_2024_betelgeuse_lsp} argue that its $\sim$2100 day LSP is best explained by a low-mass companion, as the observed light–RV phase relation and position on the PL sequence match a dusty, interacting binary. Similarly, \citet{decin_2025_binary} recently showed that companions spending part of their orbit inside the dust-formation zone can reproduce the observed semi-regular variable observability, mid-IR secondary minima,  RV curves favor the argument of periastron  $>180$ degrees, and RV–light phase lag near -90 degrees.
\\\\
Conversely, oscillatory convective modes have been predicted to arise in nonadiabatic conditions within the outermost convective zone of evolved stars \citep{cox_1980_stellar_pulsations, wood_2007_convection_pulsation_red_giants, saio_2015_oscillatory_convective_modes,wood_2015_pulsations}. It has been theoretically shown that particular dipole versions of these modes can stably exist and satisfy the observed PL relations of sequence D stars \citep{saio_2015_oscillatory_convective_modes}. Observations have revealed an overabundance of LSP stars in regions of the PL plane where primary pulsations are transitioning, suggesting a connection between LSPs and internal pulsations \citep{Trabucchi_2017_new_interp_PL}. Recent hydrodynamic simulations and synthetic stellar population models of the Magellanic Clouds further link the transition from small-amplitude red giants to semi-regular variables with a shift from stochastic driving to self-excited pulsations, coinciding with increased mass-loss rates, dust formation, and the onset of LSPs \citep{trabucchi_2025_instability_strip}. These transitions are found to depend on luminosity, effective temperature, metallicity, hydrogen content, and mixing length parameters \citep{trabucchi_2025_instability_strip}. Authors such as \citet{takayama_2020_LSP_thermal_dipole} and \citet{spaeth_2024_non-radial_oscillations} have modeled oscillatory convective modes using spherical harmonics, showing that they can reproduce LSP light curve amplitudes and periods. Fitted parameters typically indicate mean temperature variations of a few hundred Kelvin \citep{takayama_2020_LSP_thermal_dipole}. Recent 3D hydrodynamic simulations \citep{ahmad_2023_3d_sim_agb} also report the emergence of oscillatory dipole modes, albeit with periods significantly shorter than those of observed LSPs. However, these models omit key physical ingredients, such as non-gray radiative transfer and uncertainties in mixing length prescriptions, to which the properties of oscillatory convective modes are critically sensitive \citep{saio_2015_oscillatory_convective_modes}.
\\\\
Understanding the physical cause of LSPs is vital for constraining mass loss and galactic chemical enrichment. In the oscillatory convective scenario, asymmetric temperature profiles influence opacity, wind driving, and planetary nebula formation \citep{hofner_2018_mass_loss_review}. Under the binary hypothesis, companions alter mass outflow morphology and rates \citep{decin_2020_companion_wind_shaping} and challenge the “brown dwarf desert” \citep{Grether_2006_how_dry_brown_dwarf_desert}, with suggestions that some companions may be former planets that accreted mass during red giant branch (RGB) or AGB phases \citep{Nicholls_2010_ellipsoidal_variability_sED,soszyski_2021_binary_origin_LSP}. 
\\\\
To distinguish between the binary and oscillatory convective mode hypotheses, we have made high spatial resolution observations of the nearby LSP variable red giant RT Pav. By comparing these observations with parametric models and reconstructing wavelength-dependent images, in addition to Gaia constraints on any hypothetical companion separation, we aim to gain new insights into the origin of RT Pav’s LSP.
\\\\
RT Pav is an oxygen rich semi-regular variable (SRb) at a Gaia measured distance of 505$\pm$17 pc, with a primary pulsation period of 85 days and spectral type M4/5III \cite{houk_sptype_1975mcts,gaia_2020yCat.1350....0G}. RT Pav is classified as a sequence D star with a clear mid-infrared excess and a LSP in the visible light curve of 757 days \citep{payne1952variable, Houk_1963_first_LSP}. The peak-to-peak amplitude of the LSP in the V-band varies between periods, but is typically between 0.5 - 1.4 magnitude, corresponding to the flux dropping by up to a factor of 27\% of its peak value.  
RT Pav has been classified as having oxygen-rich dust emission with structured silicate features, also exhibiting a mysterious emission feature
at 13$\mu$m that is commonly seen in SRb variables \citep{sloan1996_13umsources}. Various parametric dust shell models have been fitted to infrared spectroscopic indicating effective stellar temperatures between 2700-3300K, inner and outer dust shell radii of $\sim$40 and $\sim$300mas respectively, with inner-dust temperatures around 400-500K \citep{rt_pav_disk_hashimoto_1994, hron1997semiregular, rt_pav_dust_shell_sloan_1998, ramdani2003low}. Historical high-resolution spectroscopic monitoring revealed RV variations of 3–5 km\,s$^{-1}$ \citep{feast1972kinematics, olivier_2003_origin_LSP}, leading to the hypothesis that LSPs could arise from binarity. More recently, Gaia DR3 has provided high-precision astrometry and sparse RV measurements\citep{gaia_dr3_summary_2022, gaia_nss_2022}. Although Gaia DR3 classifies  RT Pav as a single star (\texttt{non\_single\_star} flag = 0), this does not conclusively rule out the presence of a companion. The Gaia non-single star (NSS) pipeline attempts to detect binarity via deviations from single-star astrometric solutions, RV variability, or eclipsing signatures \citep{gaia_nss_2022}. For stars with high-quality astrometric fits, such as RT Pav, which has a renormalized unit weight error (RUWE) of 0.825 and an astrometric excess noise of 0.47mas, the absence of a binary classification implies no evidence for detectable orbital motion above Gaia’s sensitivity limits \citep{lindegren2021}. This will be explored further in Sect. \ref{sec:constrain_orbital_param}.

\begin{figure}[h]
    \centering 
    \includegraphics[width=8cm]{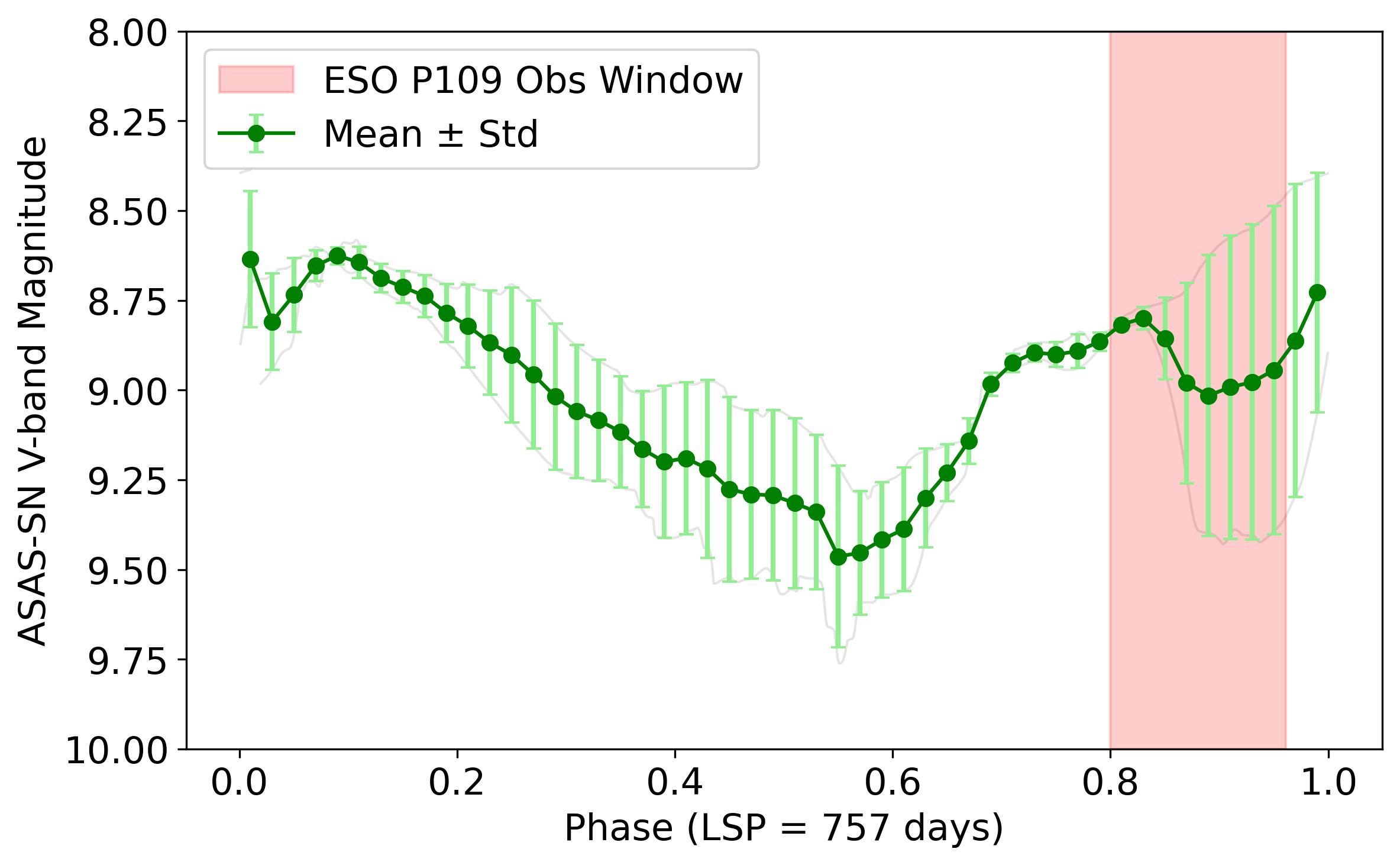}
    \caption{ Folded light curve of RT Pav from V-band ASAS photometry\protect\footnotemark{}.The green curve shows the detrended and phase-bin averaged curve over multiple LSP epochs. }
    \label{fig:RT_pav_ASAS_lightcurve}
\end{figure}
\footnotetext{ASAS photometry available here: \url{https://asas-sn.osu.edu}}

\section{Observations and data reduction}
Observations of RT Pav were conducted during ESO period 109 (1 April - 30 September 2022) at the VLTI on Cerro Paranal, Chile. At this time, RT Pav's LSP was at phase $\sim$0.8 (Fig. \ref{fig:RT_pav_ASAS_lightcurve}), when under a binary hypothesis, the companion would have the widest projected separation. Data were obtained with the 1.8\,m Auxiliary Telescopes (ATs) using three instruments: PIONIER \citep{lebouquib_2011_PIONIER_instrument}, GRAVITY \citep{grav_colab_2017_GRAVITY_instrument}, and MATISSE \citep{lopez_2022_MATISSE}. All instruments combine four telescopes: PIONIER in the H band (1.5-1.8\,$\mu$m), GRAVITY in the K band (1.9-2.4\,$\mu$m), and MATISSE in the L (3.2-3.9\,$\mu$m), and M band (4.5-5.0\,$\mu$m). Appendix~\ref{ap:vlti_obs} lists the observation details and calibrators.

\subsection{VLTI observations}
\begin{figure*}[h]
    \centering
    \includegraphics[width=16cm]{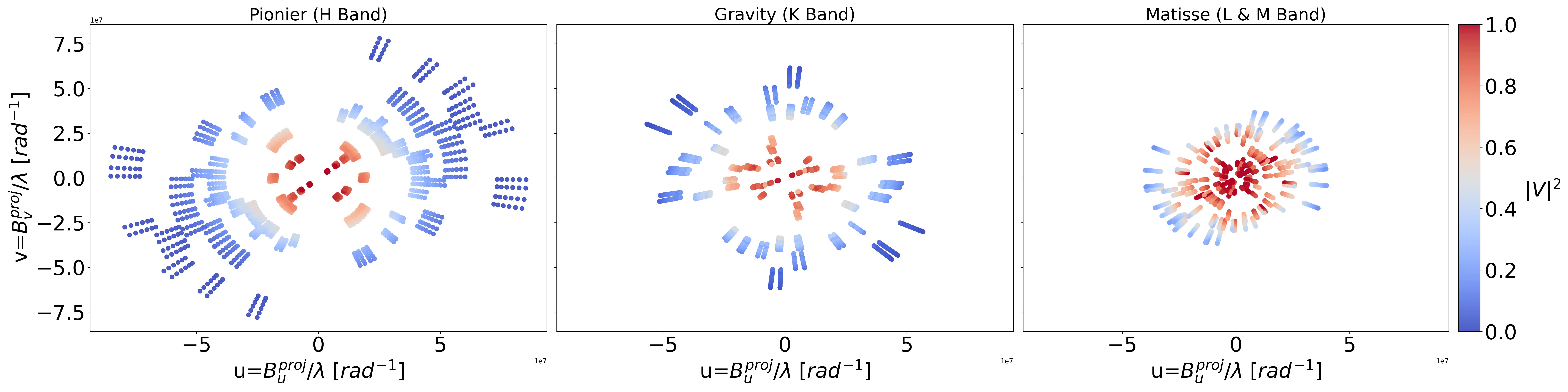}
    \caption{$(u,v)$ plane coverage of RT Pav observations on the VLTI for each instrument. Points are colored based on the square visibility measured at the respective $(u,v)$ coordinate}. 
    \label{fig:UV_coverage}
\end{figure*}

For RT Pav, we conducted an imaging program with PIONIER consisting of 14 CAL-SCI-CAL sequences, alternating calibrator (CAL) and science (SCI) observations. These sequences spanned ESO-defined small, medium, and large baselines (10-130\,m; see Figure~\ref{fig:UV_coverage}) using calibrators HD~171042, HD~172211, and HD~181019. Observations were acquired in GRISM mode, dispersing the H band light (R$\sim$40).
\\\\
A snapshot program with GRAVITY involved 4 SCI-CAL observations covering the small, medium, large, and astrometric configurations. GRAVITY observations were performed on-axis in split polarization mode at R$\sim$4000.
\\\\
MATISSE imaging comprised 12 CAL-SCI-CAL sequences. N-band data was ultimately discarded due to poor calibrators, correlated fluxes below limits that allow absolute calibrations of observables and poor data quality. For the L and M bands, MATISSE employed the SiPhot photometric mode at low resolution (R$\sim$34), measuring photometry simultaneously with fringes. Figure~\ref{fig:UV_coverage} shows the $(u,v)$ coverage and measured visibilities.

\subsection{Data reduction}
For PIONIER we used the calibrated product from the Optical Interferometry Database (OiDB) \citep{haubois_2014_OiDB_reduction}  with the data reduction processing done by IPAG with the PNDRS pipeline \citep{lebouquib_2011_PIONIER_instrument}. This contained calibrated square visibilities and closure phases. GRAVITY data reduction was performed with the GRAVITY Instrument Pipeline 1.5.0 with the default visibility calibration recipe \texttt{vis\_cal} which provides calibrated visibilities, bispectrum, and differential visibilities. Two data products observed sequentially on 2022-06-24T07:22:53 and 2022-06-24T07:35:09 suffered from oscillations in wavelength dependent quantities which was attributed to instrumental effects, and were therefore discarded. \\\\
MATISSE data were reduced using version 2.0.0 of the ESO data reduction pipeline with assistance from the Jean-Marie Mariotti Center (JMMC) expertise center. In many cases there was considerable variability between the visibilities measured on different beam commuting devices which led to large uncertainties in the merged data product. Data with uncertainties on the square visibility larger than 0.4 or closure phase larger than 20 degrees were ultimately discarded. Observables measured for different input channels (via the beam commuting device) which alternates across the observations to remove instrumental biases, were then merged into different bins for data obtained during chopping and non-chopping sequences. In general there was considerable discrepancy between the chopped and non-chopped data products which was outside the typical $\sim$0.1 uncertainty on the square visibility. This is likely due to thermal background, and therefore only quantities observed during chopping sequences were considered for the L and M bands. Three files were ultimately discarded from the L and M band observations due to large inconsistencies between beam commuted observations which are outlined in Appendix \ref{ap:vlti_obs}.

\section{Constraining the parameter space for binary versus convective mode hypotheses} \label{sec:constrain_orbital_param}
Before presenting the fitted parametric models, we first perform some simple calculations and Monte Carlo (MC) simulations to better understand, and constrain parameter space expected under the binary versus convective mode hypotheses. 

\subsection{Angular separation constraints from Gaia reflex motion and the long secondary period} \label{sec:gaia_constraints}
\begin{figure}[h]
\centering
\includegraphics[width=\columnwidth]{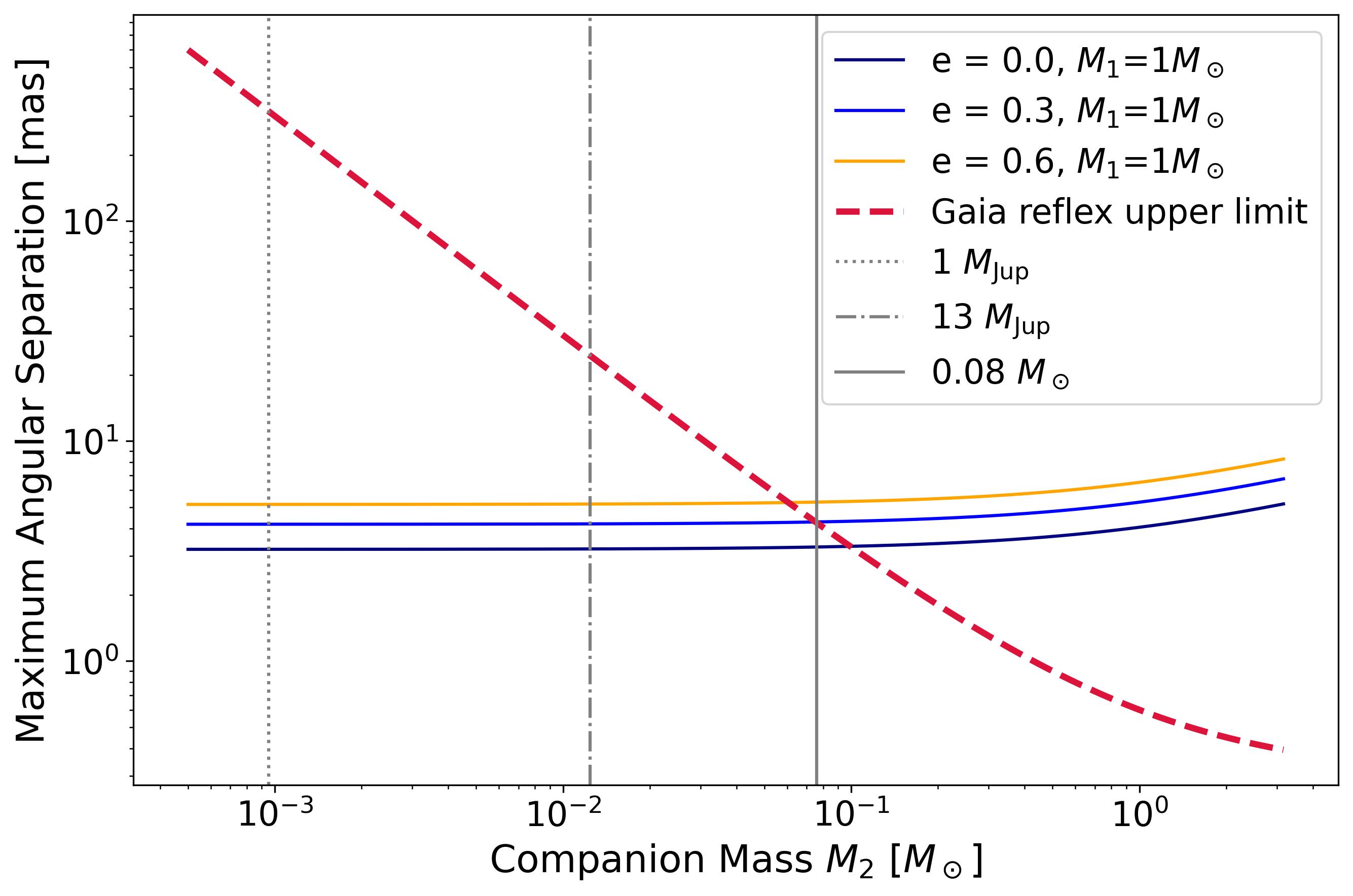}
\caption{Predicted maximum angular separation of a companion to RT Pav as a function of companion mass under the binary hypothesis assuming RT Pav's 757day LSP, and a primary mass of 1$M_\odot$. Curves are shown for orbital eccentricities of 0.0, 0.3, and 0.6. The thick dashed line marks the adopted Gaia DR3 astrometric reflex motion detection threshold of 0.3\,mas. Companion configurations above this threshold would typically induce a detectable astrometric signal and are therefore excluded by Gaia's classification of RT Pav as a single star.}
\label{fig:mas_distribution_montecarlo}
\end{figure}

If RT Pav’s 757-day LSP is due to a binary companion, then the absence of a Gaia DR3 non-single star classification implies that any reflex motion induced by such a companion lies below Gaia’s detection threshold. For bright sources such as RT Pav ($G \sim 6$, \citet{gaia_2020yCat.1350....0G}), a conservative estimate for the detectable photocenter displacement in Gaia DR3 is $a_0 \lesssim 0.3$ mas (see Figure 7 in \citet{lindegren2021}). This threshold enables us to place upper limits on the angular displacement of the primary star induced by a companion.
Assuming an orbital period of $P = 757$ days and a primary mass of $M_1 = 1.0\,M_\odot$ (appropriate for an M4/5III AGB), we compute the Keplerian semi-major axis of the binary system as
\begin{equation}
a = \left( \frac{G (M_1 + M_2) P^2}{4\pi^2} \right)^{1/3},
\end{equation}
where $M_2$ is the mass of the companion. The maximum possible projected separation is then given by
\begin{equation}
\theta = \frac{a (1 + e)}{d},
\end{equation}
with $e$ the orbital eccentricity and $d = 505$\,pc the distance to RT Pav. However, Gaia is sensitive to the motion of the photocenter, not the relative orbit. For a binary with negligible light contribution from the companion, the photocenter reflex motion of the primary is
\begin{equation}
\theta_{\text{phot}} = \frac{M_2}{M_1 + M_2} \cdot \frac{a }{d}.
\end{equation}
Imposing Gaia’s detection threshold of  $\lesssim 0.3$\,mas places an upper bound on allowed combinations of $M_2$, $e$, and projected separation. Together, the Gaia reflex motion threshold and the LSP period constraint define an allowed region of parameter space for unresolved companions, which we illustrate by computing angular separations and identifying which combinations of $M_2$ and $e$ remain consistent with Gaia’s classification of RT Pav as a single star. For a solar mass primary star, under the binary hypothesis Fig. \ref{fig:mas_distribution_montecarlo} shows valid companion separations  on the order of a 3-5 mas depending on the eccentricity of the orbit, which is close to the photosphere for a typical AGB star at RT Pav's distance. Going to the extreme mass end of an AGB of 5$M_\odot$ we get a maximum angular separation of 5-8.5mas for eccentricities 0-0.6. Considering the Gaia reflex upper limit (red curve in Fig. \ref{fig:mas_distribution_montecarlo}) the Gaia data places a upper limit on a companion mass of $~0.1M_\odot$ for companion outside a $\sim3$mas diameter of the primary.  

\subsection{Geometric bounds on a dusty companion}
\label{sec:geometric_bounds_companion}
If RT Pav’s LSP were produced by a bound companion surrounded by dust, the geometry of the system imposes strong limits on the size and location of the obscuring material.  Assuming the 757-day period represents the orbital period, the corresponding separation for a primary of $M_1=1\,M_\odot$) is $a=[G(M_1+M_2)P^2/4\pi^2]^{1/3}=1.68\,\mathrm{AU}$ for a companion mass $M_2=0.1\,M_\odot$, which represents the upper end of the mass range discussed by \citet{soszyski_2021_binary_origin_LSP} and also by our astrometric constraints (Sect. \ref{sec:gaia_constraints}). Using the Eggleton relation \citep{eggleton_1983_roche_lobe_radius}, the companion’s Roche-lobe radius is then $R_{L,2}/a=0.49q^{2/3}/[0.6q^{2/3}+\ln(1+q^{1/3})]=0.21$, giving $R_{L,2}\simeq0.35\,\mathrm{AU}$.  Independently, the visible light curve amplitude of RT Pav implies a lower bound on the projected size of any optically thick occulter: taking the maximum (single secondary cycle) observed depth of $F_{\min}/F_{\max}=0.27$ ($\Delta m=1.42$ mag) from the visible light curve (Fig. \ref{fig:RT_pav_ASAS_lightcurve}) and equating the fractional flux loss to the area ratio of a circular clump and the stellar disk, $f=1-F_{\min}/F_{\max}=0.73$ and $r_d=\sqrt{f}\,R_\star\simeq0.85\,R_\star$. For the measured UD radius of RT Pav of 0.8 $\mathrm{AU}$ (see Table \ref{tab:baseline_parameter_results} using measured distance 505\,pc), the required clump radius is $r_d\simeq0.68\,\mathrm{AU}$, exceeding the upper limit of the Roche-lobe size (via Gaia constraints) by several factors.  This discrepancy means that a dense, opaque cloud gravitationally bound to the heaviest possible companion that is not detected by Gaia, could not reproduce the optical modulation unless one or more assumptions are violated: either the obscuring material extends beyond the Roche lobe (for example through wind–Roche-lobe overflow, spiral density wakes, or circumbinary structures), or the variability arises from noncentral, partially transparent, or multiple clumps that together intercept a large fraction of the stellar light.  The lower bound on $r_d$ therefore provides a robust geometric constraint: any dusty structure capable of producing the observed photometric depth must subtend a projected area at least $70\%$ of the stellar disk, implying that the obscuring material is either large‐scale and not completely gravitationally confined or associated with mechanisms other than a compact Roche-bound envelope.

\subsection{Limits on the closure phase versus wavelength}\label{sec:closurephase_MC} 
\begin{figure*} [h]
   \begin{center}
   \includegraphics[height=8.5cm]{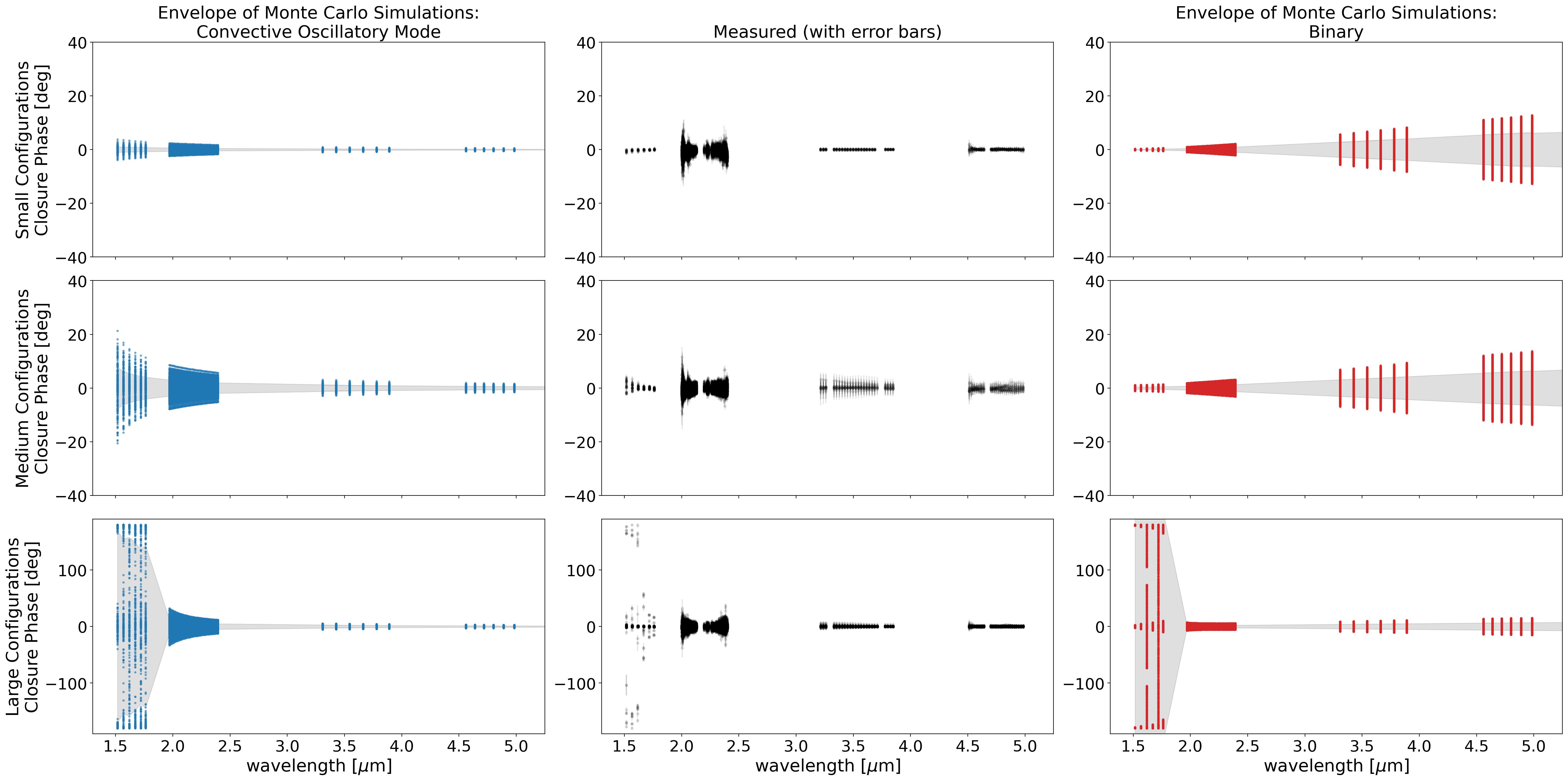}  
   \end{center}
   \caption{Measured (center column) versus Monte Carlo simulations of the closure measured by the VLTI configurations of (left) thermal variations across the photosphere with temperature gradients $\pm$ 100K, and (right) a binary enshrouded in a large dust cloud with a temperature between 1000-1500K and a radius up to 26\% the diameter of the primary. Monte Carlo simulations assume a fixed zenith target geometry without Earth rotation synthesis. The light gray shaded region highlights $\pm1\sigma$ envelope of wavelength binned closure phases across the different MC simulations. Gray colors in the center column are measurements with uncertainties. } 
   { \label{fig:mc_thermal_v_binary} }
 \end{figure*}

The geometric constraints in Sect.~\ref{sec:geometric_bounds_companion} show that the lower size limits required to reproduce the observed light-curve variation are substantially larger than the gravitationally bound volume of a companion’s Roche lobe. This motivates testing whether such large-scale structures would leave detectable interferometric signatures distinct from those expected for intrinsic stellar surface variations.  To do so, we examined the wavelength dependence of the closure phase, which provides a direct measure of departures from point symmetry in the brightness distribution. The two LSP hypotheses predict different wavelength-dependent signatures.  In the binary scenario, a cool ($1000$–$1500$\,K) dusty clump or envelope orbiting near the star introduces an additional off-axis flux component whose relative contribution increases toward longer wavelengths, producing closure phases that rise with wavelength.  In contrast, the oscillatory convective-mode hypothesis involves asymmetric thermal gradients (up to a few hundred Kelvin) across the stellar photosphere, with stronger contrast at shorter wavelengths, leading to closure phases that decrease with wavelength.  Because the closure phase represents the argument of the triple product of complex visibilities over a closed baseline triangle, it quantifies the ratio of coherent flux from symmetric versus asymmetric structures while remaining insensitive to atmospheric phase noise.  VLTI instruments routinely achieve $\sim$1° precision, corresponding to sensitivity to asymmetries at the $10^{-3}$ flux level. We simulated the expected closure-phase behavior versus wavelength using Monte Carlo methods for both scenarios.  In the binary model, the primary was represented as a 3000\,K uniform disk with its measured limb-darkened diameter, and the companion as a cool ($1000$–$1500$\,K) uniform disk corresponding dust temperature estimates of \citet{soszyski_2021_binary_origin_LSP} and the conservative lower size limit implied by the observed flux variation in the visible light curve (Sect. \ref{sec:geometric_bounds_companion}).  In the oscillatory convective-mode model, the primary exhibits low-order thermal perturbations across its disk with amplitudes up to $100$\,K consistent with observations by \citet{takayama_2020_LSP_thermal_dipole}.  For each case, random orientations and phase realizations were generated and sampled over the VLTI baseline configurations of our observations. Figure \ref{fig:mc_thermal_v_binary} compares the simulated closure-phase behavior with the measurements.  Significant nonzero closure phases are measured in the H band on the longest baselines where we resolve the second visibility lobe, indicating asymmetries on scales smaller than the stellar disk, while K band closure phases reach up to 10 degrees.  No statistically significant nonzero closure phases are observed in the L or M bands.  The overall decrease of closure-phase amplitude toward longer wavelengths and the absence of measurable asymmetry in the thermal infrared are more suggestive of the oscillatory convective hypothesis for RT Pavs LSP.
 
\section{Parametric models} 
A series of baseline parametric models, including models for the binary and oscillatory convective mode hypothesis, were fitted to the observational data of RT Pav. The best fitting parametric models were used as priors for image reconstruction. The baseline models considered were:
\begin{itemize}
    \item Uniform Disk (UD) - 1 fitted parameter per waveband
    \item Power-2 limb-darkening (LD) - 2 fitted parameters per waveband
    \item Uniform ellipse (EL) - 3 fitted parameters per waveband
    \item Unresolved binary (B) - 4 fitted parameters per waveband
    \item Oscillatory convective dipole (T) - 4 fitted parameters per waveband
\end{itemize}
Mathematical descriptions of these models can be found in the following sections. For each model, we fit the observed square visibilities and closure phases. To assess the quality of the fit we typically measured the reduced chi-square defined as: 
\begin{equation}
\begin{aligned}
\chi^2_\nu
&= \frac{1}{N_V}\sum \left( \frac{V^2_{\text{obs}} - V^2_{\text{model}}}{\sigma_{V^2}} \right)^2 \\
&\quad + \frac{2}{N_{\psi}}\sum \left( \frac{1-\cos(\psi_{\text{obs}}-\psi_{\text{model}})}{\sigma_{\psi}^2} \right)
\end{aligned}
\end{equation}
Where $V^2$ and $\psi$ are the square visibilities and closure phase, with  independent measurements $N_V$ and $N_\psi$ respectively, while $\sigma$ is the uncertainty on the measured quantity estimate. A $\chi^2$ term for closure phase was used that respects 2$\pi$ phase wraps in the difference of measured and observed values, similar to that adopted in \cite{CP_ref_2018ApJ...857...23C}. The model fits, binned within continuum wavelengths, are tabulated in Table \ref{tab:baseline_parameter_results}. In some cases fits were also performed as a function of wavelength in smaller wavelength bins to analyze the wavelength dependence. Parametric modeling and fitting made use of the  python packages \texttt{PMOIRED}\citep{PMOIRED_2022ascl.soft05001M}, \textsc{CANDID} \citet{gallenne_2015_candid}, in addition to \texttt{emcee} \citep{emcee_2013} which implements a python version of the affine-invariant ensemble sampler for Markov chain Monte Carlo (MCMC) algorithm. Given the larger number of parameters for the uniform ellipse and binary models, a grid search was typically performed prior to fitting. Sampling for the grid search was set empirically to ensure smooth chi-squared maps to minimize the risk of missing global minima.

\subsection{Uniform disk model}\label{sec:uniform_disk}
The visibility (V) for a uniform disk intensity distribution of angular diameter $\theta$ (rad) is given:
\begin{equation} \label{eq:ud_disk_visibility}
    V(\rho|\theta) = \frac{2J_1( \pi \rho \theta )}{ \pi \rho \theta } 
\end{equation}
where $J_1$ is the first order Bessel function, $\rho=\sqrt{u^2+v^2}=|B|/\lambda$ (rad$^{-1}$) is the spatial frequency that is being sampled at the point in the uv-plane given by modulus of the observed baseline length divided by the observed wavelength ($\lambda$). 

\subsection{Limb darkening model}
To measure the limb darkening of RT Pav we consider a power law limb-darkened model  \citep{hestroffer1997powerlaw_limb_darkening,limb_darken_2021A&A...654A..19D}. This was considered to capture more complex molecular layers in the stars circumstellar envelope:
\begin{align}
    V &= \frac{ \nu \Gamma(\nu) 2^{\nu} J_{\nu}(\pi \rho \theta_{LD}) }{(\pi \rho \theta_{LD})^{\nu}}
\end{align}
where $\nu = \alpha_{LD}/2 + 1$, where $\alpha_{LD}$ is the power law intensity exponent for $\mu$ - the cosine of the angle between the direction perpendicular to the stellar surface and the direction toward a distant observer. i.e,
\begin{align}
    I_\lambda(\mu)/I(\lambda(1) = \mu^\alpha
\end{align}
and where $\theta_{LD}$ is the limb darkened diameter. Limb darkened models were only fitted on PIONIER and GRAVITY data where the stellar photosphere was semi-resolved.

\subsection{Ellipse model}
To explore radial asymmetries a uniform ellipse model was considered. Analytical expressions for the visibility of an ellipse can be derived from the uniform disk model (Sect. \ref{sec:uniform_disk}) by considering properties of affine and Fourier transforms. This is outlined in Appendix \ref{ap:ellipse_model}. 

\subsection{Binary model}
The visibility of a binary system follows directly from the linearity and translation properties of the Fourier transform. It can be expressed as
\begin{equation}
V(u,v) =
\frac{V(\rho\,|\,\theta_p) + f\,G(u,v)\,V(\rho\,|\,\theta_c)}{1 + f},
\end{equation}
where $\theta_p$ and $\theta_c$ are the angular diameters of the primary and companion, $V(\rho\,|\,\theta)$ is the uniform-disk visibility from Equation~5, $f$ is the companion-to-primary flux ratio, and $G$ is the complex phase term describing the spatial offset of the secondary:
\begin{equation}
G(u, v) = \exp \big( 2\pi i\,[u\,\Delta\alpha + v\,\Delta\delta] \big),
\end{equation} 
with $\Delta\alpha$ and $\Delta\delta$ the projected right ascension and declination offsets. For the parametric models we assume the companion is unresolved, and we constrain its position to lie beyond the interferometric resolution limit. The outer search radius is chosen to encompass separations up to $\sim25$~mas, which is roughly twice the upper limit of an LSP-causing companion inferred from the Gaia astrometric constraints. Finite spectral bandwidth introduces a chromatic attenuation of the fringe contrast known as bandwidth smearing. Following \citet{gallenne_2015_candid}, this effect is modeled in \textsc{CANDID} by multiplying the companion term by a coherence factor
\begin{equation}
G(\zeta) = \left|\frac{\sin \zeta}{\zeta}\right|, \qquad 
\zeta = \pi (u\,\Delta\alpha + v\,\Delta\delta)\, / \,(R\,\lambda),
\end{equation}
where $R$ is the spectral resolution, $\lambda$ the central wavelength, and $(u,v)$ the projected spatial frequencies. This factor corresponds to the analytical integration of the complex visibility over the finite spectral bandpass and reproduces the classical loss of visibility amplitude at large separations or long baselines. We implement these constraints during the fitting procedure using \textsc{CANDID} \citep{gallenne_2015_candid}, where we jointly fit squared visibilities and closure phases over a grid of starting positions $(\Delta\alpha,\Delta\delta)$ and performs local $\chi^2$ minimizations. We also employ \textsc{CANDID}’s remove-companion and refit and detection-limit maps to evaluate the robustness of potential solutions and corresponding flux-ratio upper limits.  Complementary fits with \textsc{PMOIRED} \citep{PMOIRED_2022ascl.soft05001M} were used in an exploratory sense to confirm consistency, test wavelength-dependent behavior and generate synthetic observations.

\subsection{Oscillatory convective dipole model}\label{section:convective_modes}
Oscillatory convective dipole modes are modeled as a low order spherical harmonic \citep{wood_2007_convection_pulsation_red_giants, takayama_2020_LSP_thermal_dipole, spaeth_2024_non-radial_oscillations}, with the local temperature profile expressed in spherical coordinates:
\begin{align}
    T_{\text{eff,local}} = T_{\text{eff}} + \delta T_{\text{eff}} \operatorname{Re}\left[ \frac{Y_l^m(\theta, \phi)}{\max\big(\operatorname{Re}(Y_l^m(\theta, \phi))\big)} e^{i(2\pi \nu t + \psi_T)} \right], \label{eq:temperature_variation}
\end{align}
where:
\begin{itemize}
    \item $T_{\text{eff}}$ is the mean effective temperature.
    \item $\delta T_{\text{eff}}$ is the amplitude of temperature variation.
    \item $\psi_T$ is the phase offset of the temperature variation.
    \item $Y_l^m(\theta, \phi)$ are spherical harmonics defined as:
    \begin{align}
        Y_l^m(\theta, \phi) = \sqrt{\frac{2l+1}{4\pi} \frac{(l-m)!}{(l+m)!}} P_l^m(\cos \theta) e^{im\phi}, \label{eq:spherical_harmonics}
    \end{align}
    with $P_l^m(\cos \theta)$ being the associated Legendre polynomials.
    \item $\max\big(\text{Re}(Y_l^m(\theta, \phi))\big)$ ensures normalization.
\end{itemize}
We fixed the effective temperature at 3000\,K (mid-range from \citealt{hron1997semiregular, ramdani2003low}) and adopted a convective dipole mode (l=1, m=0) for our simulations, since $Y^{\pm1}_0$ is equivalent to a rotated $Y^1_0$ from the observers view point. This significantly simplifies the model to:
\begin{equation}
    T_{\text{eff,local}} = T_{\text{eff}} + \delta T_{\text{eff}} \cos({\theta}) \cos({2\pi \nu t + \psi_T}) 
\end{equation}
Due to degeneracy between the line of sight and oscillation phase, we set the phase to 0.8 (matching the observed LSP), leaving the stellar diameter, inclination, projection angle, and temperature variation amplitude as free parameters. In our spherical coordinate system, the dipole axis is aligned with the $z$-axis. The inclination $i$ (polar angle) and projection angle $PA$ (azimuth measured counterclockwise from the $x$-axis) define the observer’s viewpoint; $i=90^\circ$ is pole-on, while $i=0^\circ$ is edge-on. The angular intensity profile was computed by projecting the black-body image (from the 3D temperature profile) along the line of sight over wavelength bins, and sampling its Fourier transform at the observed baselines. The thermal oscillation model was fitted only to PIONIER data, where the stellar photosphere was resolved.

\section{Parametric modeling results}
\begin{figure*} [h!]
   \begin{center}
   \begin{tabular}{c c c c} 
   \includegraphics[height=3.2cm]{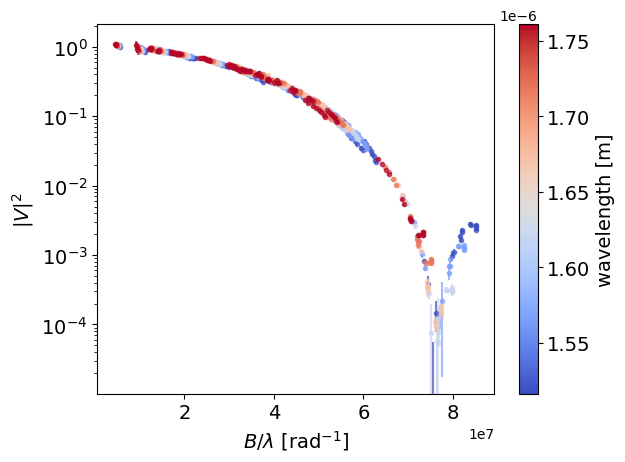} & 
   \includegraphics[height=3.2cm]{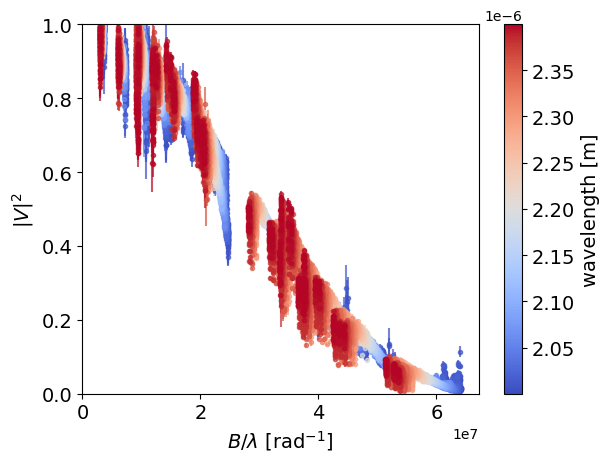} & 
   \includegraphics[height=3.2cm]{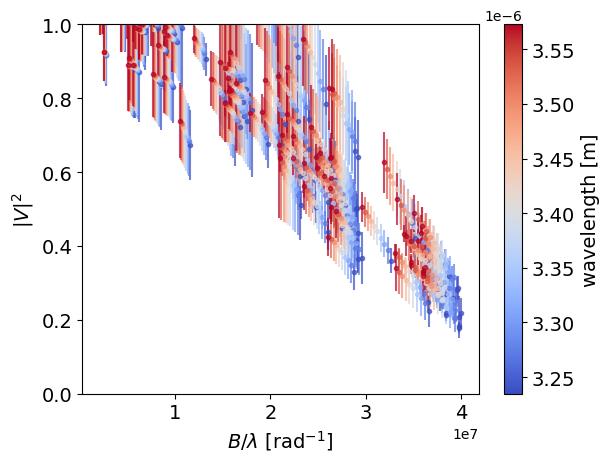} & 
   \includegraphics[height=3.2cm]{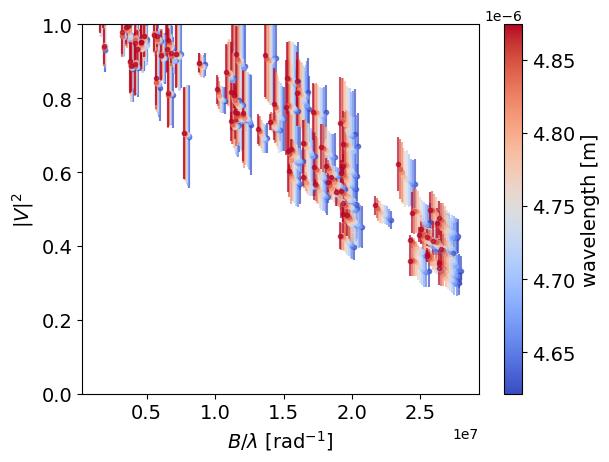}  
   \end{tabular}
   \end{center}
   \caption{ Square visibilities versus angular frequency for RT Pav on PIONIER (H band), GRAVITY (K band), MATISSE (L band), and MATISSE (M band) data.  } 

   { \label{fig:V2_all} }
 \end{figure*}

\begin{figure}[h]
    \centering
    \includegraphics[width=8cm]{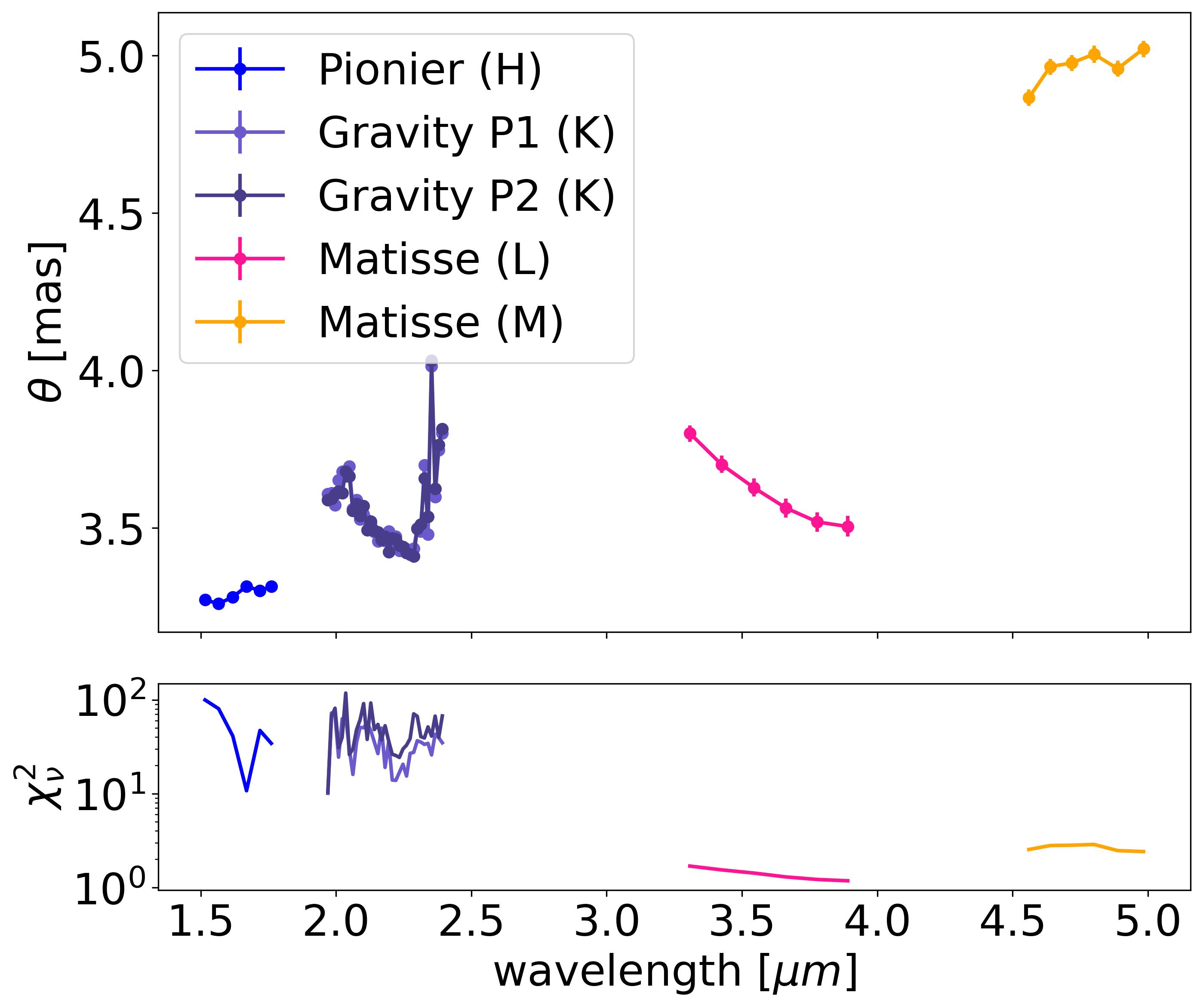}
    \caption{ Fitted uniform disk diameters for H, K, L, M bands from the PIONIER, GRAVITY and MATISSE $|V|^2$ data. Bottom plots show the reduced chi square per wavelength bin fit on the square visibility (not including closure phase). For the GRAVITY data we fitted both polarization channels separately (P1 and P2 respectively).}
    \label{fig:ud_fit}
\end{figure}

Below we summarize the results for the parametric fits per instrument. The measured square visibilities and closure phases used for the fits are shown in figures \ref{fig:mc_thermal_v_binary} and \ref{fig:V2_all}. Figure \ref{fig:ud_fit} shows the fitted (using only the square visibility) uniform disk diameters versus wavelength, while continuum waveband binned results can be found in Table \ref{tab:baseline_parameter_results}.
 
\begin{table*}
\caption{\centering Best-fit parametric models}
\centering 
\begin{tabular}[h]{||c || c | c c c c ||}
 \hline
 Model & Fitted & H band & K band & L band & M band \\ [0.5ex] 
  & Parameters &  &  &  &   \\ [0.5ex] 
 \hline\hline 
 \multirow{2}{*}{Uniform disk} 

    & $\theta_{UD}$ [mas] & 3.311 $\pm$ 0.009 & 3.481 $\pm$ 0.024 & 3.675 $\pm$ 0.011 & 4.974 $\pm$ 0.010 \\  \cline{2-6} 
    & $\chi^2_{\nu, UD}$ & 91.3 & 18.7 & 1.9 & 2.7  \\
  \hline\hline
 \multirow{3}{*}{Limb darkened} 
    & $\theta_{LD}$ [mas]  & 4.191$\pm$0.021 & 4.277$\pm$0.157 & - & - \\ \cline{2-6}
    & $\alpha_{LD}$ & 1.727 $\pm$ 0.011 & 1.757$\pm$ 0.411 & - & -  \\ \cline{2-6}
     & $\chi^2_{\nu, LD}$ & 59.3& 14.8 & - & -  \\

 \hline\hline 
 \multirow{4}{*}{Uniform ellipse} 
     & $\theta_{EF}$ [mas] & 3.364 $\pm$ 0.005 & 3.540 $\pm$ 0.023 & 3.811 $\pm$ 0.027 & 5.333 $\pm$ 0.022  \\ \cline{2-6}
     &  incl. [deg] & 17.29 $\pm$0.39 & 12.540 $\pm$ 0.096 & 18.92 $\pm$ 1.590 & -26.42 $\pm$ 0.720   \\ \cline{2-6}
     & proj. angle [deg] & 22.787 $\pm$ 0.033 & 80.6 $\pm$ 0.82  & 128.190 $\pm$ 4.570  & -32.900 $\pm$ 1.250 \\ \cline{2-6}
     & $\chi^2_{\nu, EF}$  & 55.1 & 12.5  & 1.9 & 2.4 \\ 
 \hline\hline 
 \multirow{5}{*}{Binary} 
     & $\theta_{UD,B}$ [mas] & 3.296 $\pm$ 0.002 & 3.4709 $\pm$ 0.010  & 3.623 $\pm$ 0.066 & 4.911 $\pm$ 0.040 \\ \cline{2-6}
     &  $\Delta$RA(x) [mas]  &  3.014$\pm$0.019 & 4.690 $\pm$ 0.175 & -2.090 $\pm$ 1.650 & 2.793 $\pm$ 1.690 \\  \cline{2-6}
     & $\Delta$DEC(y) [mas] & 0.463 $\pm$ 0.039 & -11.093 $\pm$ 0.512 & 4.301 $\pm$ 3.00 &  -6.710 $\pm$ 1.70 \\ \cline{2-6}
     & $F_c/F_*$ [\%] & 1.007 $\pm$ 0.048 &  0.546  $\pm$ + 0.337  & 1.580 $\pm$ 0.70  &  3.660 $\pm$ 0.720 \\ \cline{2-6}
     & $\chi^2_{\nu, B}$  &  15.0 & 14.9 & 0.2 & 0.3 \\ 
      \hline\hline 
\multirow{5}{*}{Convective Dipole} 
     & $\theta_{UD,T}$ [mas] & 3.3 $\pm$ 0.03 & -  & - & - \\ \cline{2-6}
     &   incl. [deg]  & 170 $\pm$ 49  & -  & - & -  \\  \cline{2-6}
     & proj. angle [deg]  & 172 $\pm$ 38 & - & -  & -  \\ \cline{2-6}
     & $\delta T$ [K]   & 212 $\pm$ 141 & - & -  & -    \\ \cline{2-6}
     & $\chi^2_{\nu, T}$  &  30.0 &   &  &   \\ 
 \hline 
 \hline 
\end{tabular}
\tablefoot{Best-fit parametric models for each instrument. Models were fitted in wavelength binned continuum bands. All uncertainties provided are purely statistical and do not account for calibration errors or uncertainties related to wavelength calibration.}
\label{tab:baseline_parameter_results}
\end{table*}

\subsection{PIONIER - H band}
PIONIER is the only dataset that properly resolves the stellar photosphere. Using \texttt{CANDID}, we find a best fit with $\chi^2_\nu\simeq15.0$ at $\Delta{\rm RA}=2.418\pm0.025\,{\rm mas}$, $\Delta{\rm DEC}=0.040\pm0.100\,{\rm mas}$, $f=1.65\pm0.14\%$, and $\theta_{\rm UD}=3.298\pm0.005\,{\rm mas}$ (719 degrees of freedom, with formal  $n\sigma\approx34$ if spectral channels are treated as independent).
When this best-fit binary is analytically removed and the data refitted, a weaker aliasing lobe remains at $(\Delta{\rm RA},\Delta{\rm DEC})\approx(-0.66,-2.45)\,{\rm mas}$ with $f\simeq0.54\%$ and $\chi^2_\nu\simeq10.4$ (formal $n\sigma\approx7.3$). To quantify false positives from sampling and noise, we repeated the search on synthetic datasets. For a uniform-disk model  sampled at the exact $(u,v)$ points with the measured uncertainties, spurious companions appear at $|\Delta{\rm RA}|\sim3\,{\rm mas}$, $f\simeq0.2\%$ with $n\sigma\approx6.5$, but the significance collapses to $\sim1.2\sigma$ after subtracting that fit. For synthetic data generated from the best-fit oscillatory dipole photosphere, a similar $\chi^2$ minimum appears along the west–east axis ($f\simeq0.1\%$, $n\sigma\approx2$), again dropping to $\sim1\sigma$ after removal.
\\
Finally, dividing the observations by the oscillatory convective dipole model in the complex visibility space (so $V^2_{\rm res}=V^2_{\rm obs}/V^2_{\rm dip}$ and ${\rm CP}_{\rm res}={\rm CP}_{\rm obs}-{\rm CP}_{\rm dip}$) yields no coherent binary detection in the residuals above the 3$\sigma$ flux detection limit. Injection–recovery tests on the observed PIONIER data after companion removal indicate a $3\sigma$ sensitivity of $\Delta m_H\simeq5.5\pm0.5$ (i.e.,\ $f\lesssim0.4\%$) beyond $\sim2$mas. Grid fits of the synthetic and real observations are shown in Appendix \ref{ap:binary_grid_search} while Fig. \ref{fig:candid_fitmap_on_dipole_residuals_pionier} shows the binary fit map on the residuals of the best-fit oscillatory convective dipole model. 
\\\\
In parallel, an oscillatory convective dipole on the photosphere provides a physically motivated alternative with $\chi^2_\nu\simeq30$; the best fit indicates a dipole axis inclined by $\sim$10 degrees from west-east, with the peak temperature at the PIONIER epoch located toward the west and a temperature amplitude of order $200$ K, consistent with typical values reported by \citet{takayama_2020_LSP_thermal_dipole}. We imposed no prior on the temperature amplitude, and although an oscillatory dipole introduces a wavelength-dependent photocenter shift, simulations for RT Pav show that even strong ($\sim$200 K) temperature dipoles generate peak-to-valley astrometric excursions of $\sim15 - 135\mu$as, which is below the effective Gaia sensitivity required to trigger a non-single-star solution. Figure \ref{fig:PIONIER_thermal_dipole_fit} shows the MCMC posteriors and best-fit image, Fig. \ref{fig:RT_pav_ASAS_lightcurve_convective_mode_fit} illustrates that the fitted phase and amplitude reproduce the visible LSP modulation, and Fig. \ref{fig:PIONIER_v2_cp_thermal_dipole_fit} compares the model $V^2$ and CP to the data. A closer look at the visibilities near angular frequencies $\sim60\,{\rm M}\lambda$ (where the closure phases are small) shows a $\sim1\%$ deficit relative to photosphere-only models, corresponding to $\sim3\%$ near-symmetric flux that could arise from a molecular layer or inner dust scattering.
\\\\
Quantitatively, the H band binary fit achieves a lower $\chi^2_\nu$ than our simple dipole parameterization ($\sim15$ vs.\ $\sim30$), and also is at a separation that is consistent with the LSP under Gaia constraints (Fig. \ref{fig:mas_distribution_montecarlo}). However, care should be taken in interpreting this as a genuine detection given the anisotropic $(u,v)$ sampling, which produces a persistent west–east minimum in the $\chi^2$ map, and the complete disappearance of any coherent companion signal in the photospheric dipole-removed residuals. Partial Fourier coverage can cause first-order photospheric dipole asymmetries to alias as off-axis point sources in parametric binary fits (see Appendix \ref{ap:binary_grid_search}), further cautioning against over interpretation of localized $\chi^2$ minima. A true companion should yield consistent $(\Delta\mathrm{RA},\Delta\mathrm{DEC})$ across independent datasets and wavebands (H, K, L, and M) and exhibit a physically plausible spectral flux ratio. The absence of such cross-band coherence would make this binary interpretation tentative.
\begin{figure}[h]
    \centering
    \includegraphics[width=8cm]{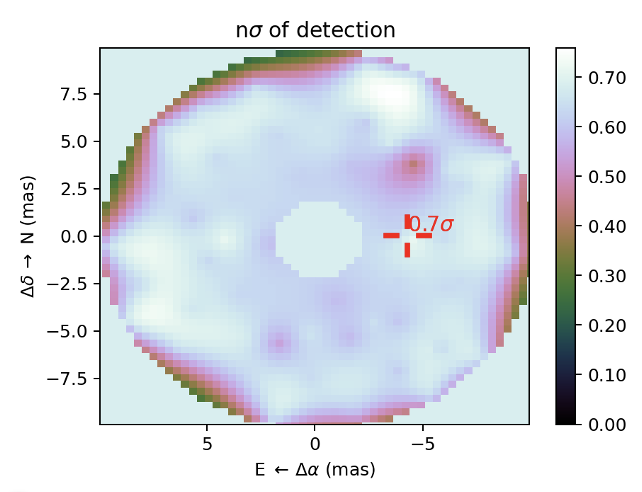}
    \caption{Fit map of a binary model on the residuals of the best-fit oscillatory convective dipole model when removed from the observed data. No significant binary signals are found above the 3$\sigma$ detection limits.}
    \label{fig:candid_fitmap_on_dipole_residuals_pionier}
\end{figure}
\begin{figure}[h]
    \centering
    \includegraphics[width=8cm]{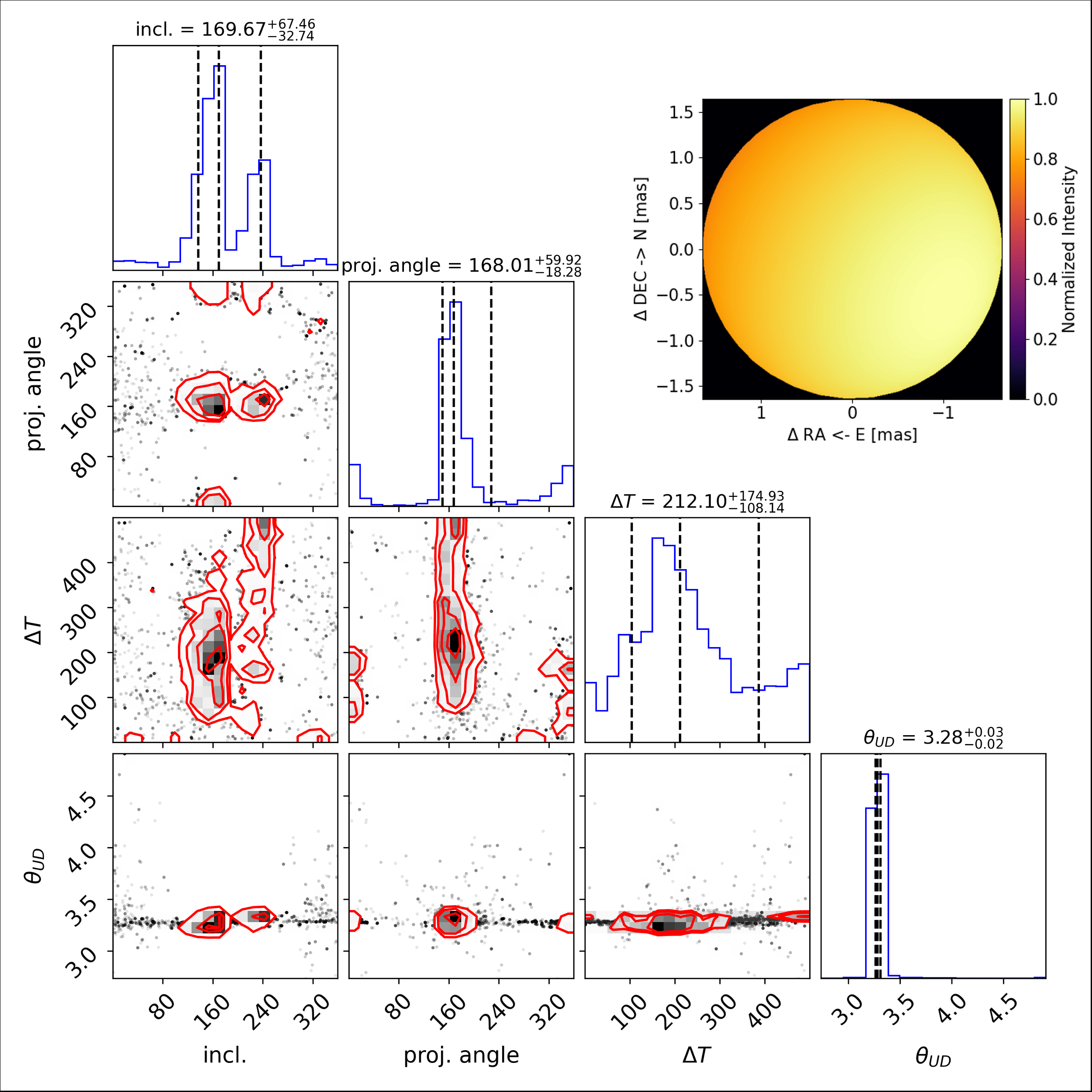}
    \caption{Left: Corner plot of a MCMC fit of the convective dipole model with PIONIER data. Right: Image of the best-fit convective dipole model during P109. }
    \label{fig:PIONIER_thermal_dipole_fit}
\end{figure}
\begin{figure}[h]
    \centering 
    \includegraphics[width=8cm]{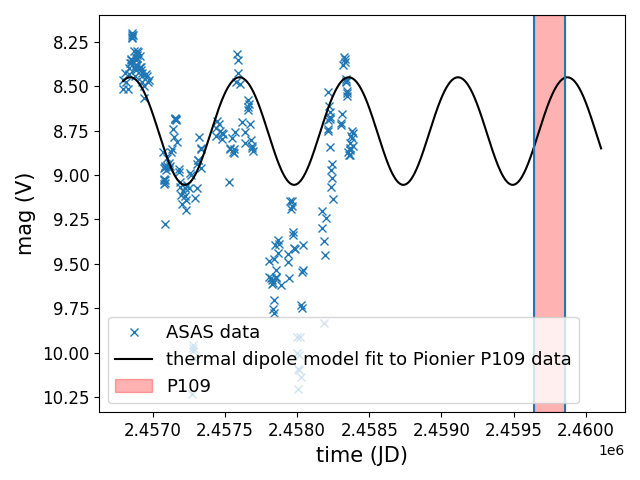}
    \caption{ Light curve of RT Pav from ASAS photometry\protect\footnotemark{} along with the independent best-fit oscillatory convective dipole model.} 
    \label{fig:RT_pav_ASAS_lightcurve_convective_mode_fit}
\end{figure}
\begin{figure*}[h]
    \centering
    \includegraphics[width=8cm]{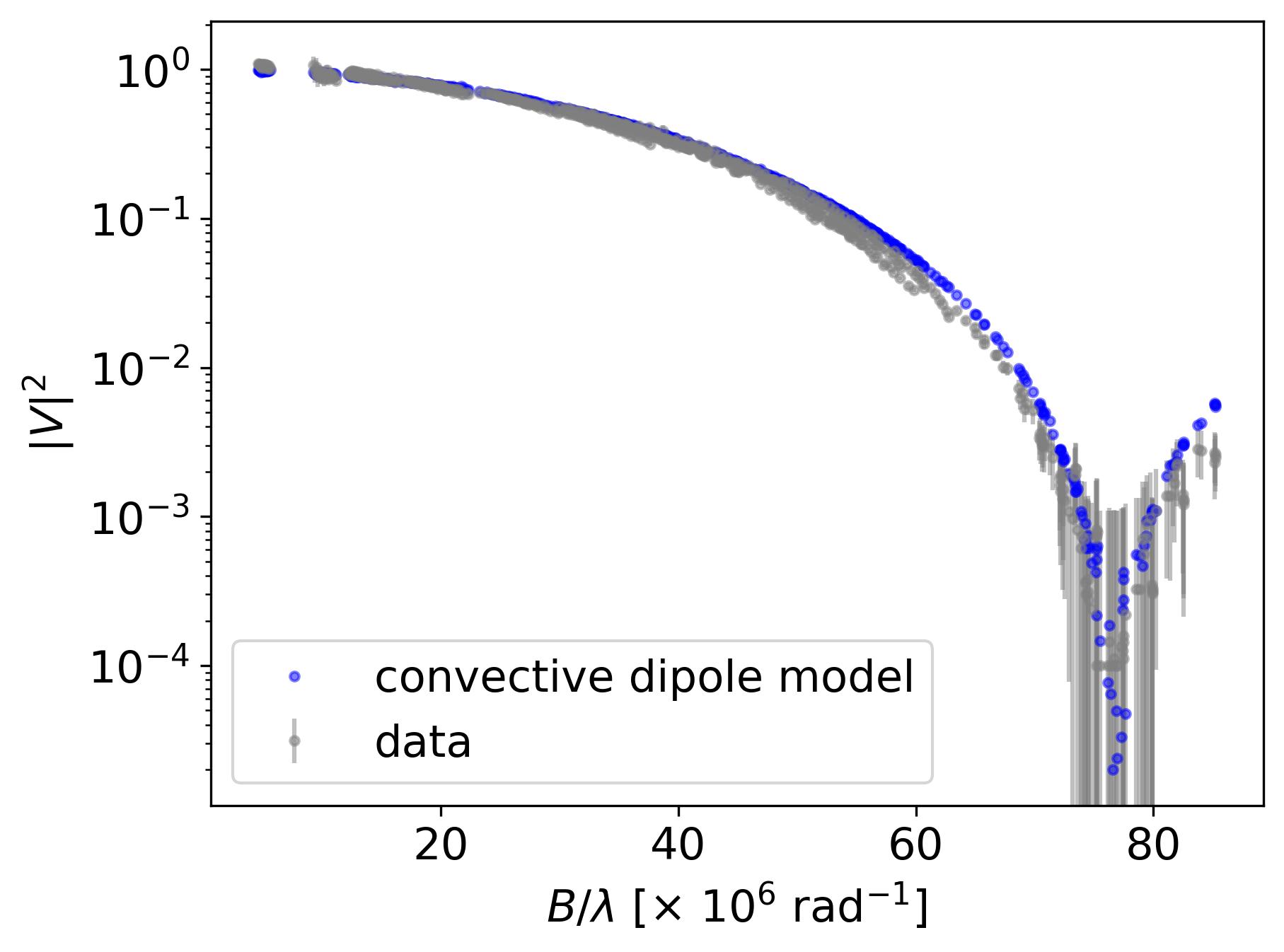}
    \includegraphics[width=8cm]{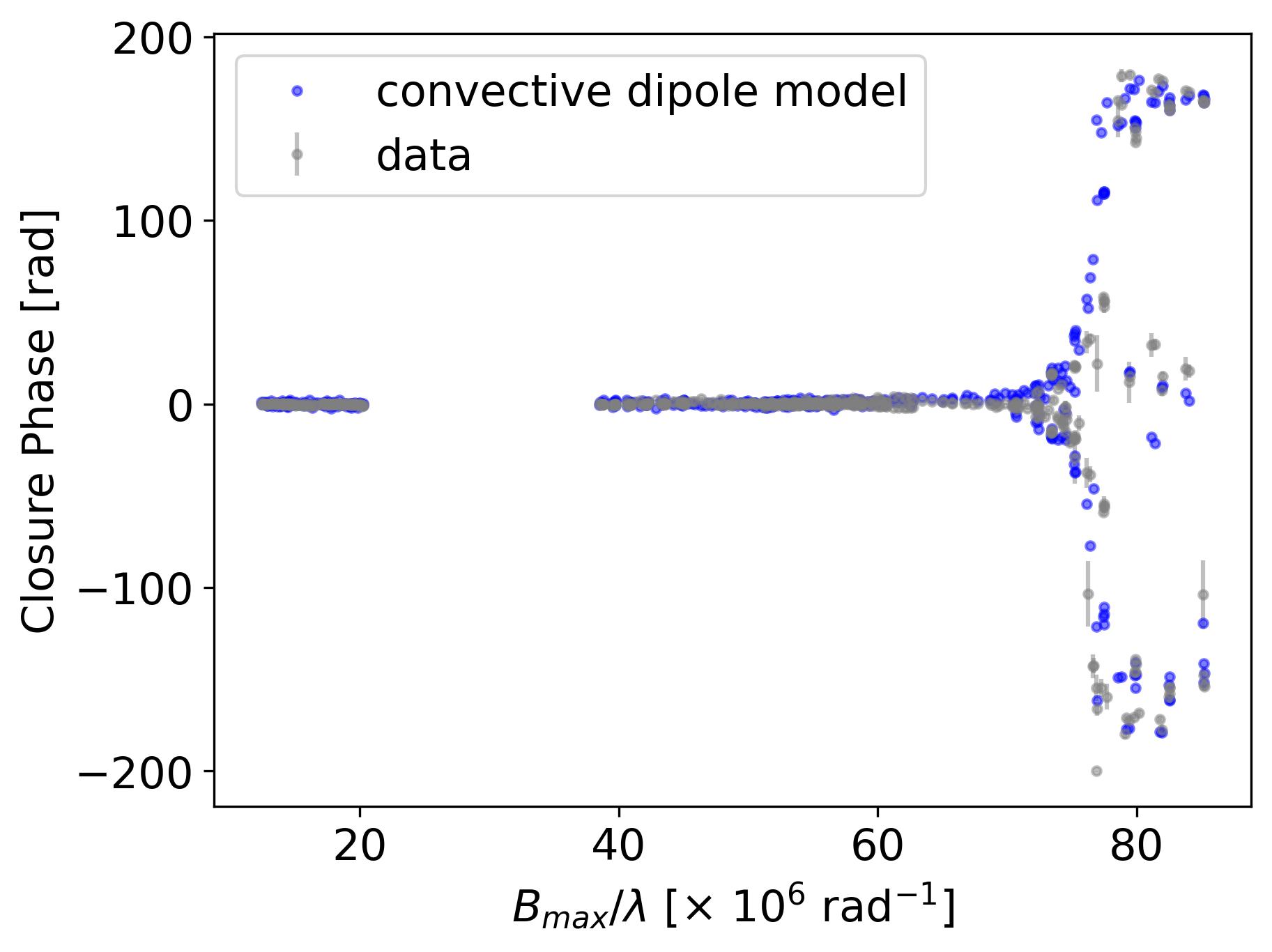}
    \caption{ Comparison of the fitted square visibilities and closure phases of the convective dipole model to the PIONIER  data.}
    \label{fig:PIONIER_v2_cp_thermal_dipole_fit}
\end{figure*}

\subsection{GRAVITY - K band}
\begin{figure}[h]
    \centering
    \includegraphics[width=7cm]{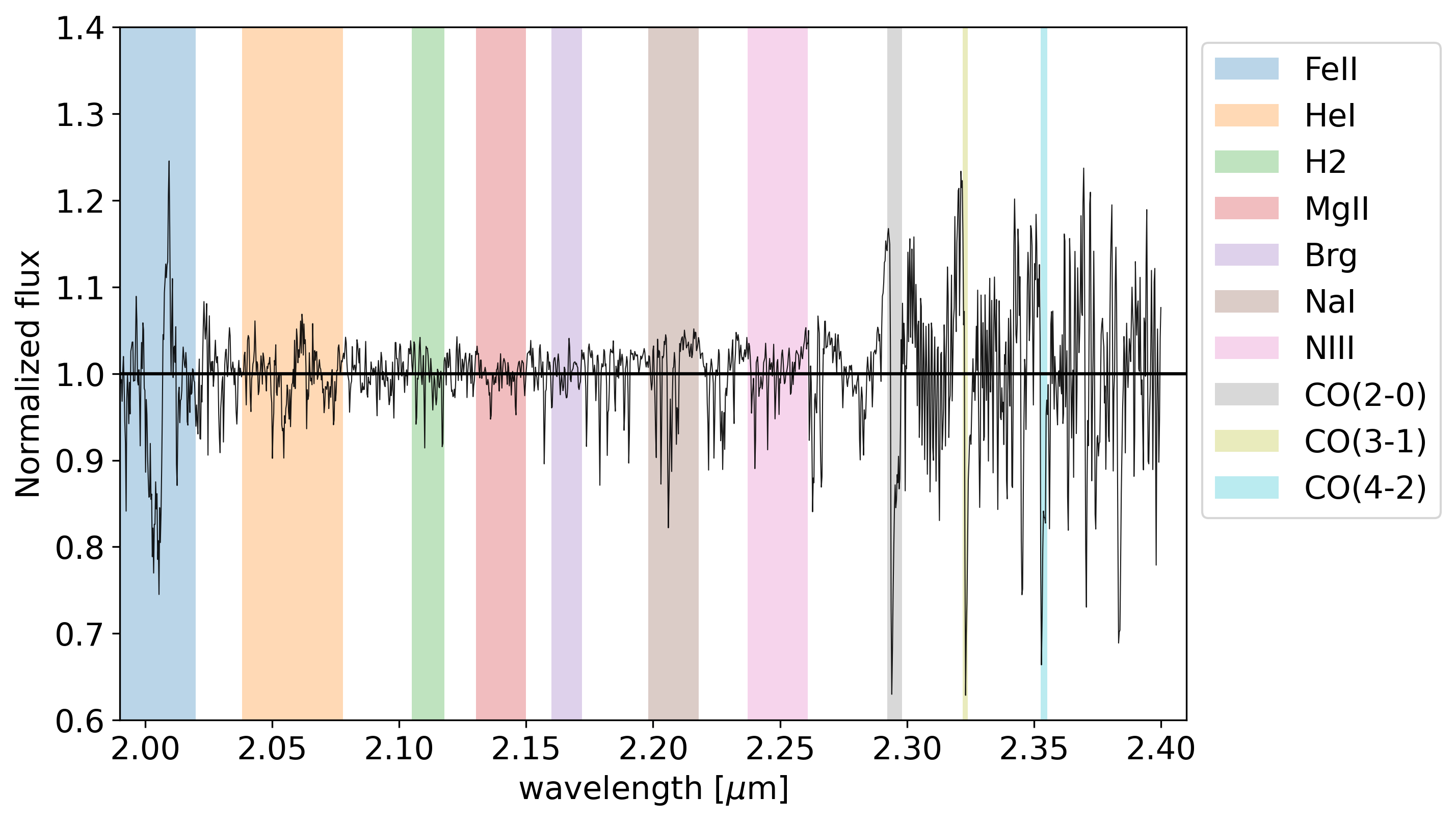}
    \caption{ GRAVITY normalized flux after the telluric correction highlighting known spectral features. }
    \label{fig:GRAVITY_spectrum}
\end{figure}

The best fitting baseline model for RT Pav in the K band is a uniform ellipse with an inclination and projection angle corresponding to an eccentricity $\sim$ 20\% and a semi-major axis aligned -10 degrees from the western direction. It is interesting that this also corresponds to the orientation of the fitted dipole axis for the oscillatory convective dipole mode in H band. 
Unfortunately, the GRAVITY data did not formally resolve the uniform disk diameter or have a highly significant convective dipole model, and is therefore not presented in Table \ref{tab:baseline_parameter_results}. The best fitting binary model (which achieved a reduced chi-squared only a few worse than the ellipse model, and similar to the limb-darkened model) fitted a ($1\sigma$ level) companion at $\Delta$RA,$\Delta$DEC = -4.69$\pm$0.18, -11.09$\pm$0.51 which is a different orientation and much wider than the fitted companion from the PIONIER data. It is also at a separation well outside the Gaia constraints outlined in Sect. \ref{sec:gaia_constraints}. Furthermore the best-fit flux ratio of 0.546$\pm$0.337\% was very close to lower 3$\sigma$ detection limit of 5.6 magnitude (0.57\%). 
Unlike the other interferometric instruments used in this study, GRAVITY provides simultaneous spectro-interferometric capabilities with high spectral resolution ($R\sim4000$). This allows us to probe line-by-line visibility and closure-phase variations across diagnostic features that trace accretion, and molecular layers.  After telluric correction and normalization of GRAVITY's flux spectrum, Fig. \ref{fig:GRAVITY_spectrum} shows various spectral features including a weak emission line for HeI at 2.06$\mu$m, and absorption lines for FeII at 2.05$\mu$m and the CO band heads between 2.29-3.6$\mu$m. No Brackett-$\gamma$ emission is detected in the GRAVITY spectrum. This rules out the presence of strong accretion shocks involving ionized hydrogen ($\gtrsim 20$ km s$^{-1}$), but does not exclude the possibility of lower-velocity, cooler accretion flows. Thus, if accretion is occurring, it must proceed without producing detectable Brackett-$\gamma$ emission, for example, through a disk partially filling the Roche lobe rather than via strong shocks. This constrains the accreting companion hypothesis of \cite{soszyski_2021_binary_origin_LSP}. We did not observe any signature in the visibility or closure phase for any spectral features besides the CO bandheads which have a significant decrease in the square visibility and increase in the closure phase for up to 10 degrees, indicating that we resolve an extended, asymmetric MOLsphere \citep{tsuji_2000_molsphere}. The uniform disk fits within these absorption regions give diameters of 3.90$\pm$0.05, 4.06$\pm$0.07, 3.94$\pm$0.06mas at wavelengths 2.294, 2.322, 2.353$\mu$m, corresponding to CO(2-1), CO(3-1), CO(4-2) band heads respectively. 

\subsection{MATISSE - L/M band}
The L and M band datasets have substantially larger uncertainties than the H and K bands, and do not strongly discriminate between simple models. The uniform disk models the data adequately with reduced chi-square 1.9 and 2.7 for L and M bands respectively, and  we see an increasing UD diameter between the L-M bands from 3.6-4.9mas. The corresponding closure phases are consistent with zero within the quoted errors (Fig. \ref{fig:mc_thermal_v_binary}), as expected for nearly centro-symmetric structure. Power-law limb darkening did not converge robustly for MATISSE because the fits tended to negative exponents (implying unphysical Gamma-function evaluations); more fundamentally, credible limb-darkening constraints require sampling near or beyond the first visibility null (second lobe), whereas the L/M baselines probe only part of the first lobe. We therefore omit MATISSE limb-darkening results from Table \ref{tab:baseline_parameter_results}. Formally, the binary models achieved the lowest reduced chi-square values of all parametric fits; however, values below unity indicate that the model fits the data more closely than expected from the stated uncertainties, possibly reflecting overestimated errors or mild overfitting. The best-fit binary positions were found at different separations, on opposite sides of the sky with $\Delta \text{RA} = -2.0, 2.8$mas, $\Delta \text{DEC} = 4.3,-6.7$mas for L and M bands respectively, and with an increasing flux ratio with respect to wavelength from 1.6\% to 3.7\%. These flux ratios were only slightly above the estimated 3$\sigma$ detection limit of 4-5 mag (Depending on the specific injection method adopted) at the given separation and were detected with a 0.9 and 11.5$\sigma$ level relative to a UD model. This higher contrast companion is expected in the M band where, under the binary hypothesis, we begin to resolve cooler dust features. However here, similar issues to the H band data apply. In particular, we find nearly symmetric $\chi^2$ minima about the field center—consistent with lower-sensitivity regions arising from anisotropic $(u,v)$ sampling, and no companion is recovered at nearby positions in other bands. Accordingly, caution is warranted in interpreting the M band result as a robust detection.

\section{Image reconstruction}
\begin{figure*} [h!]
   \begin{center}
   \begin{tabular}{c c c c} 
   \includegraphics[height=3.2cm]{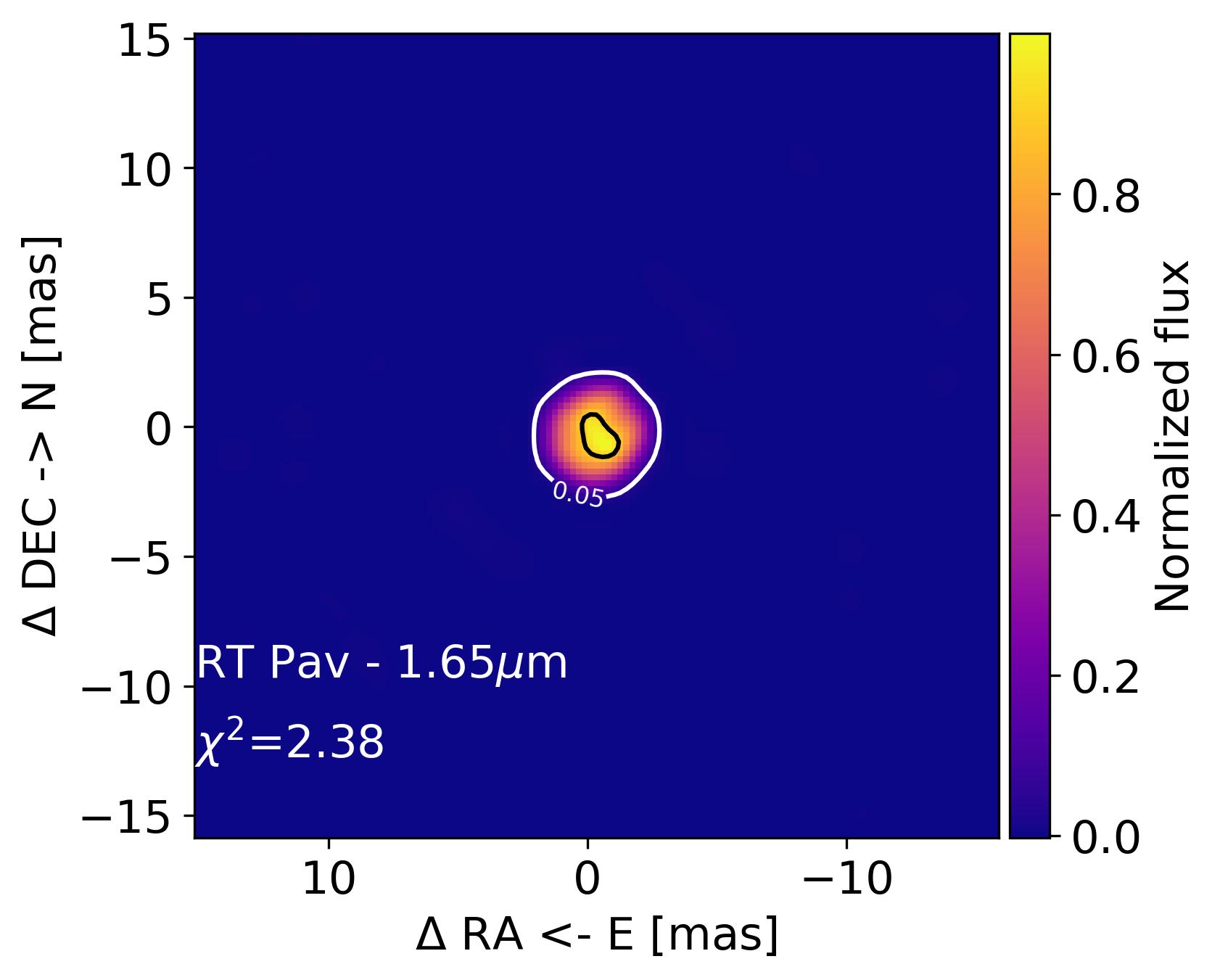} & 
   \includegraphics[height=3.2cm]{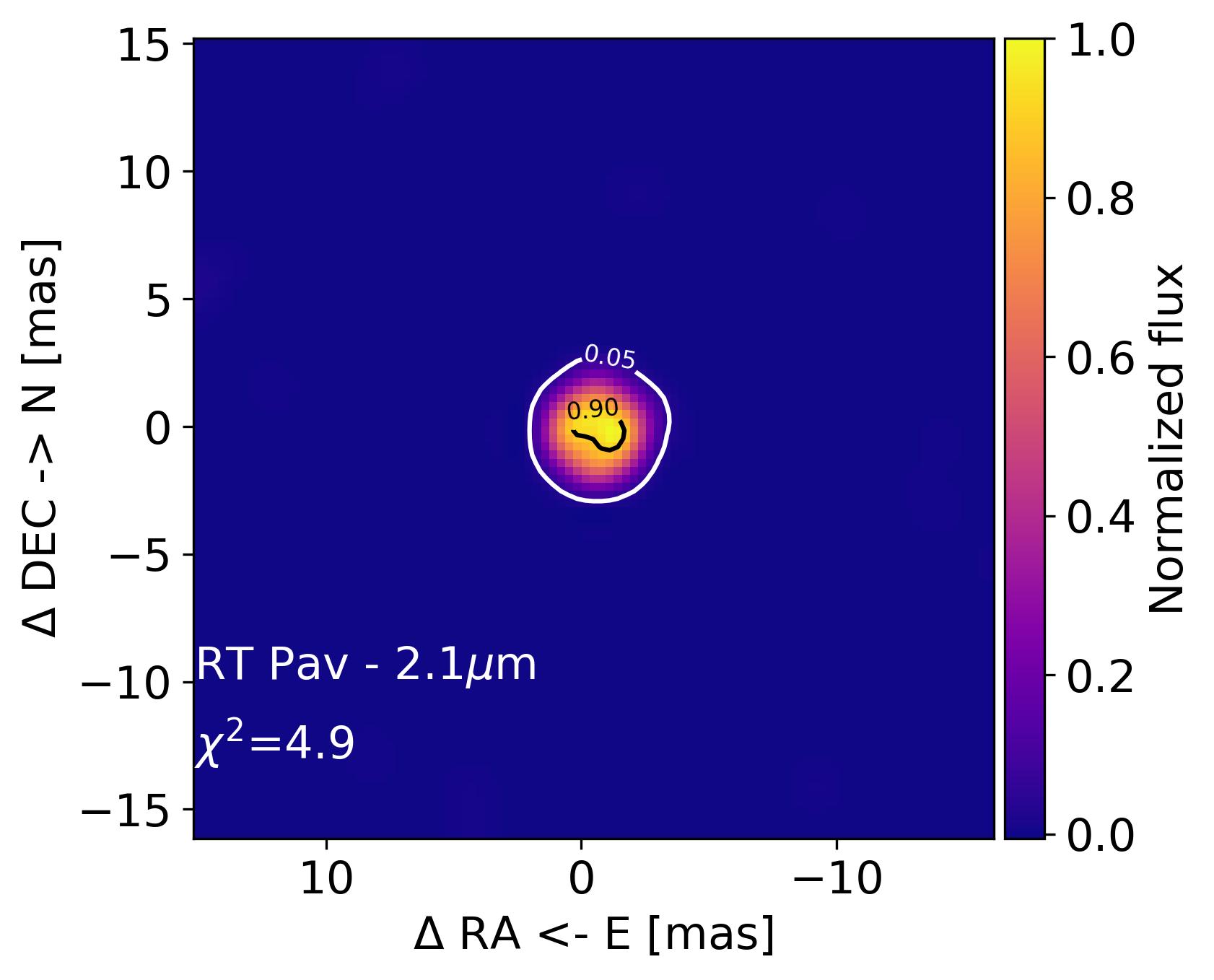} & 
   \includegraphics[height=3.2cm]{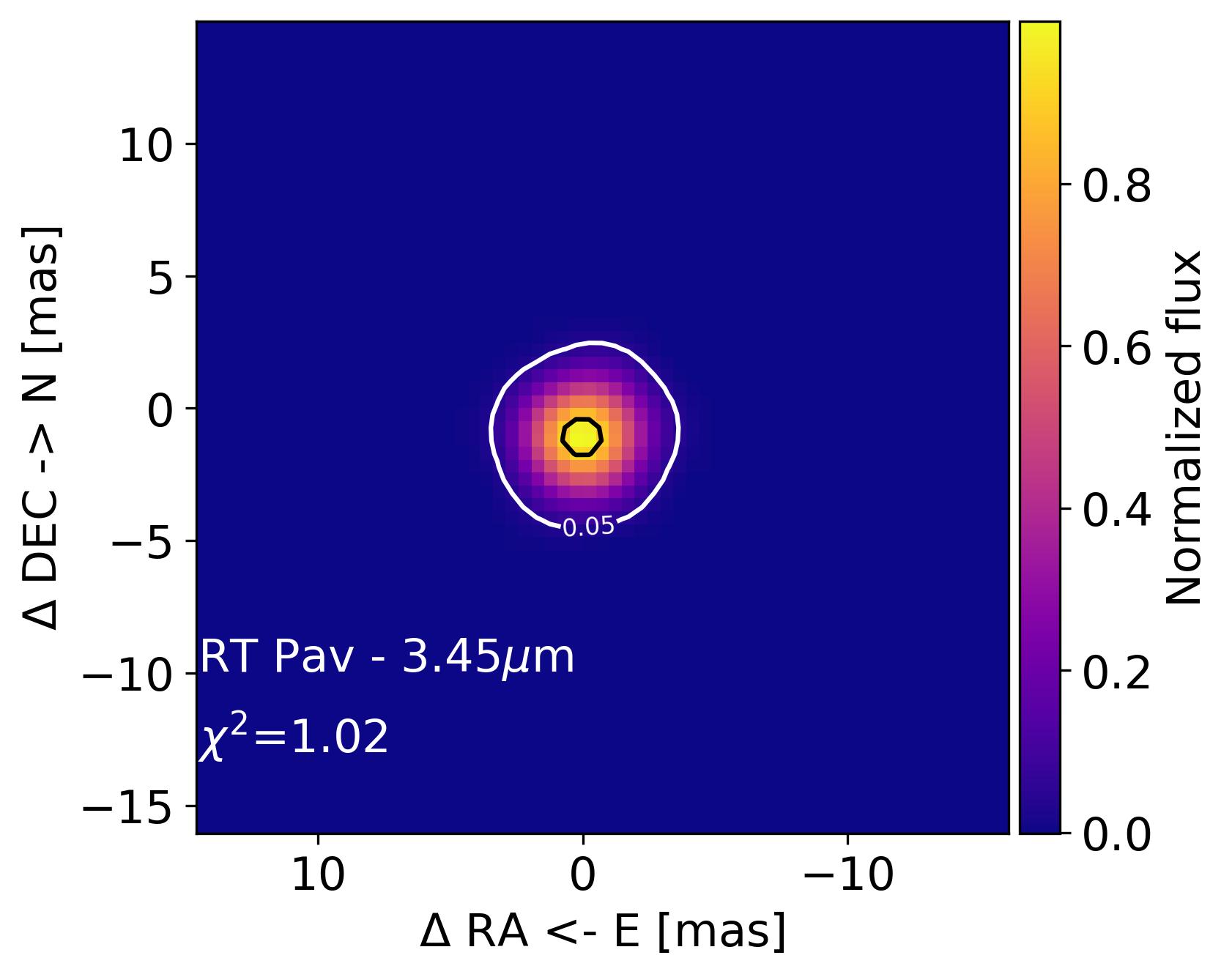} & 
   \includegraphics[height=3.2cm]{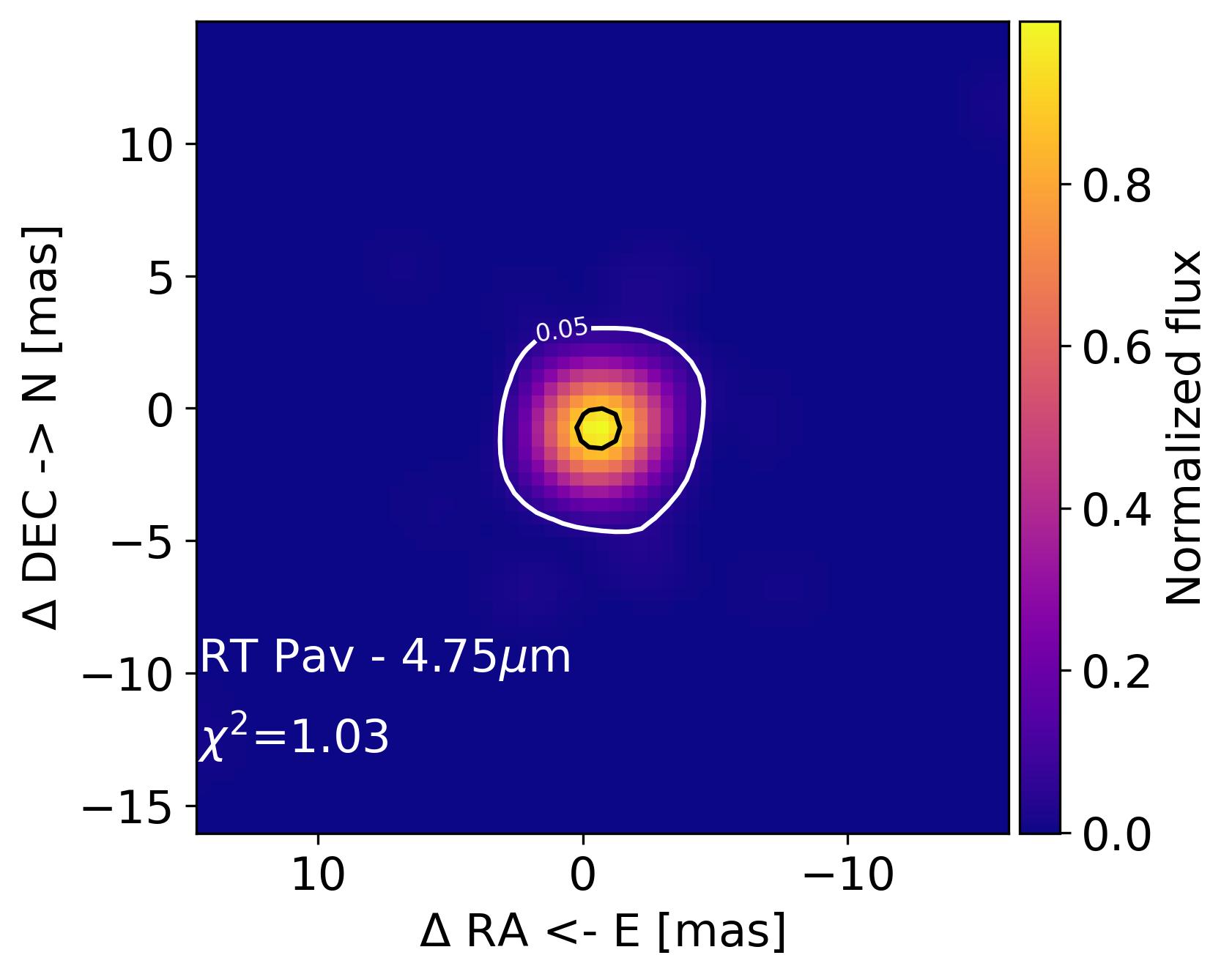}  \\
   \includegraphics[height=3.2cm]{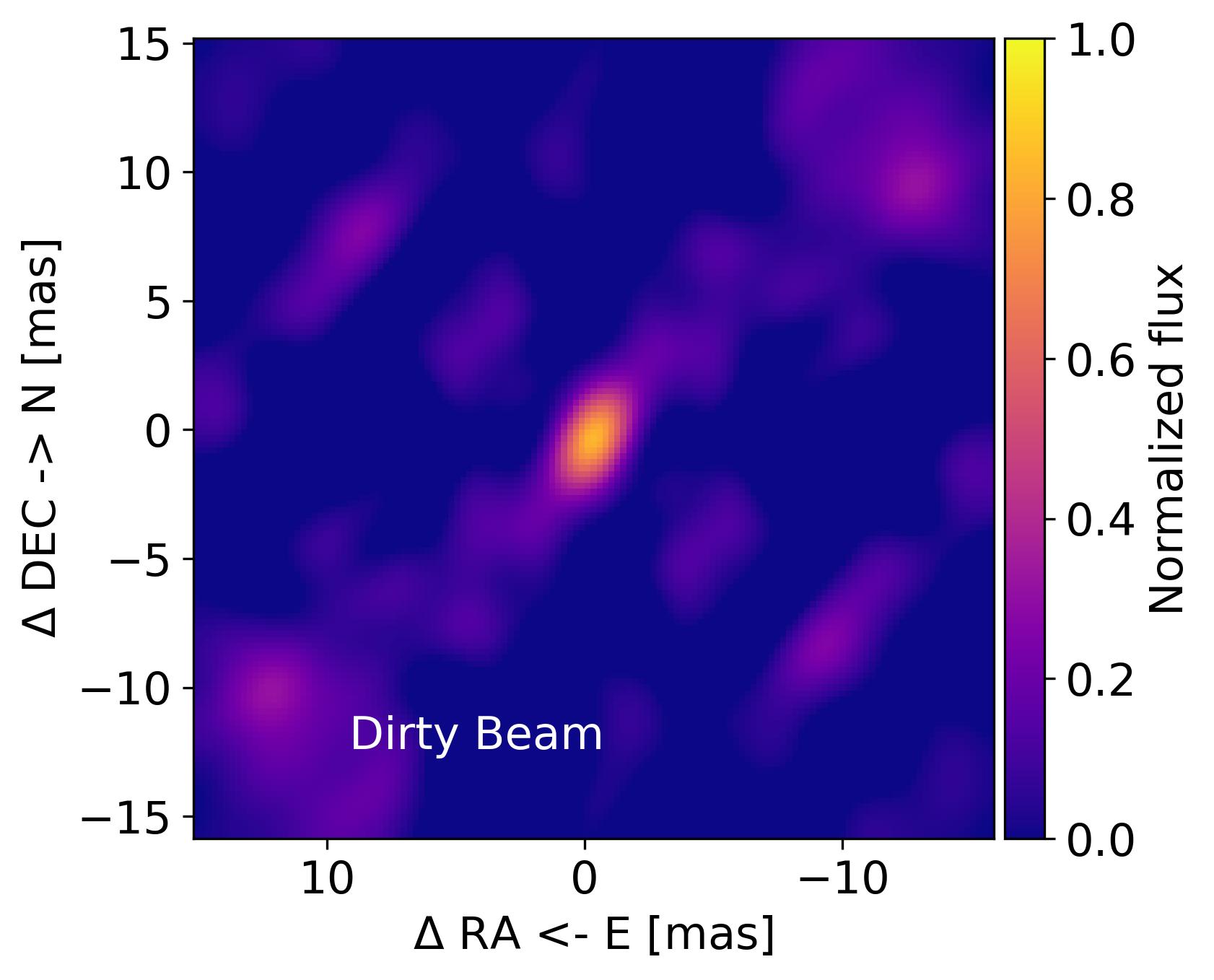} & 
   \includegraphics[height=3.2cm]{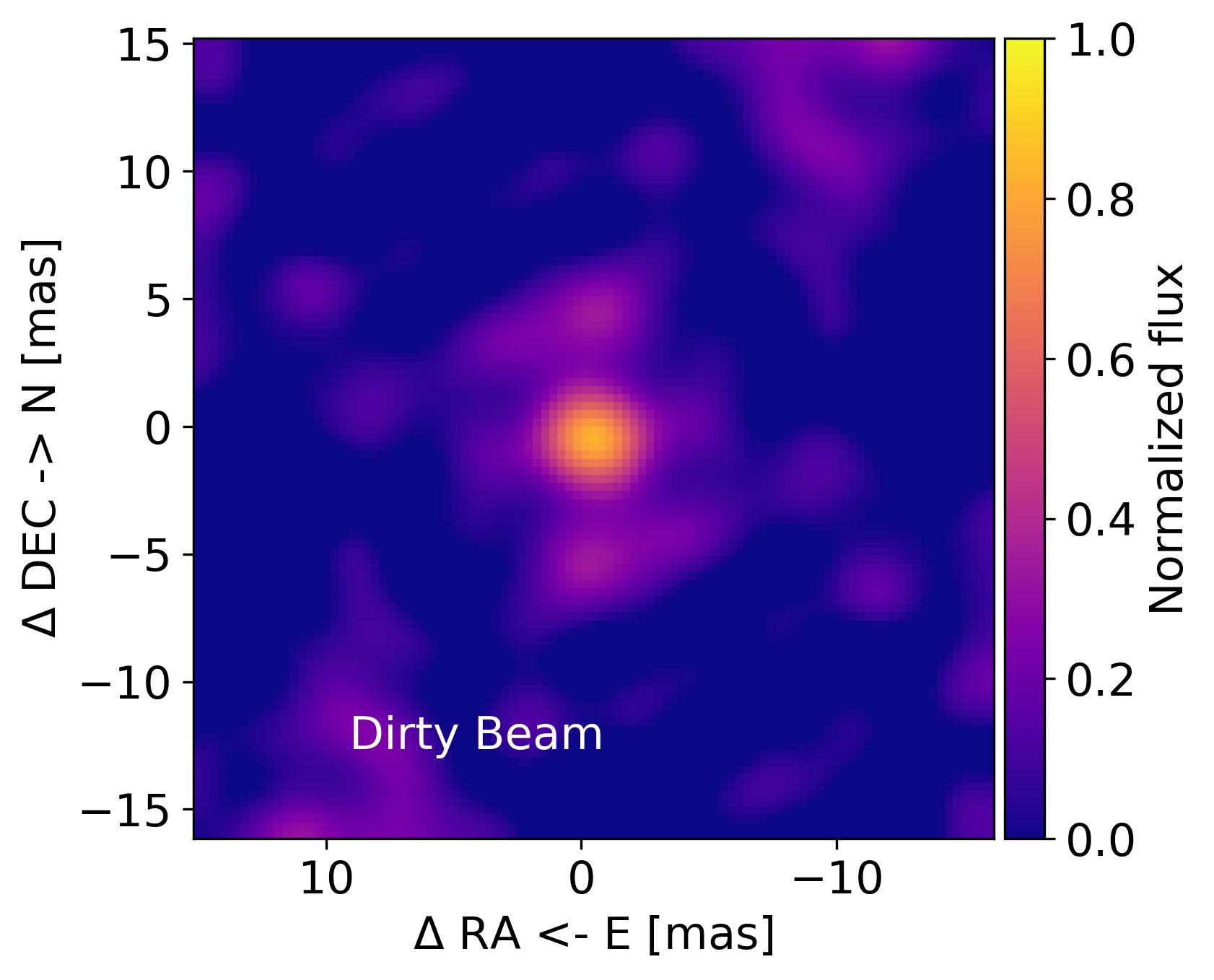} &
   \includegraphics[height=3.2cm]{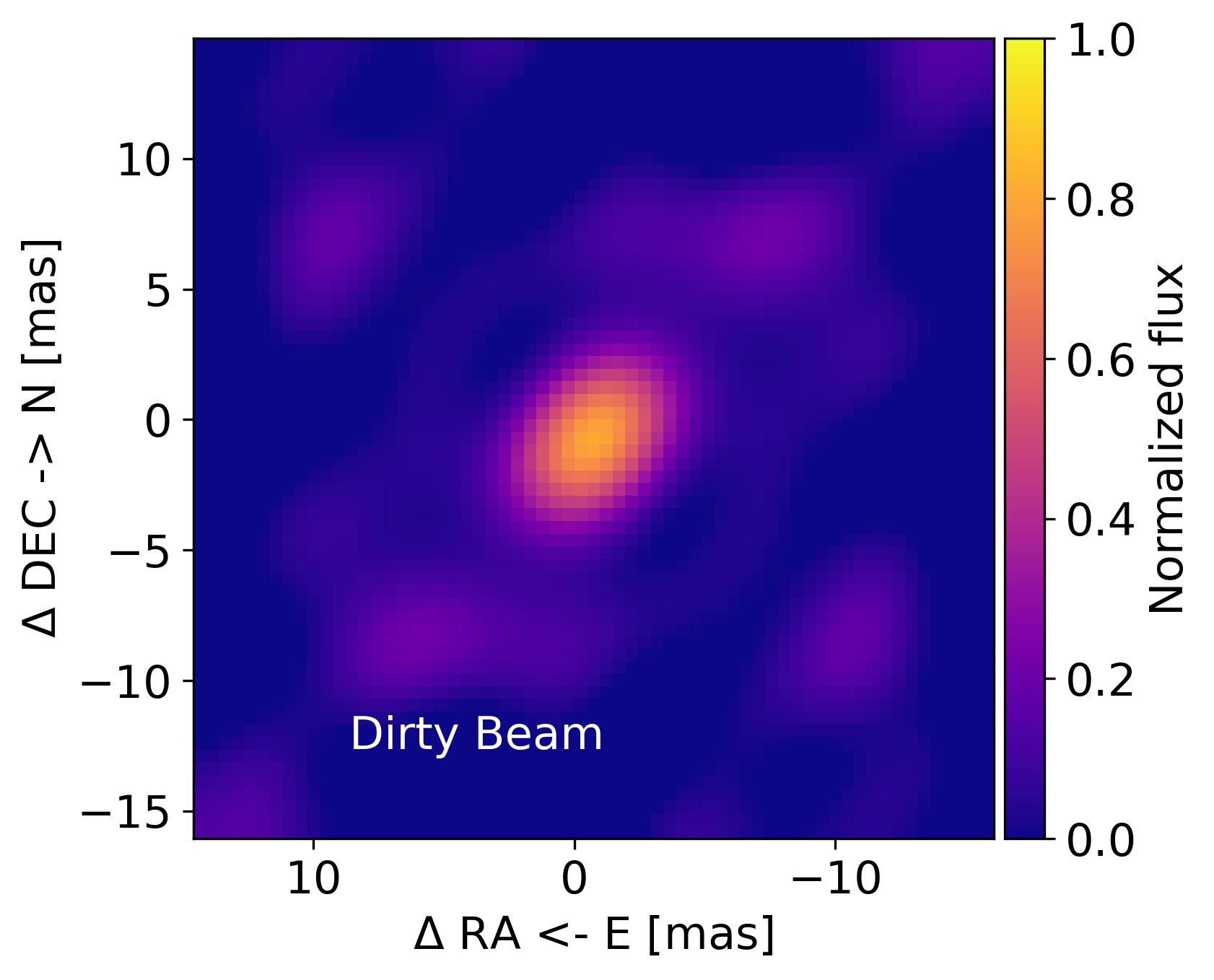} &
   \includegraphics[height=3.2cm]{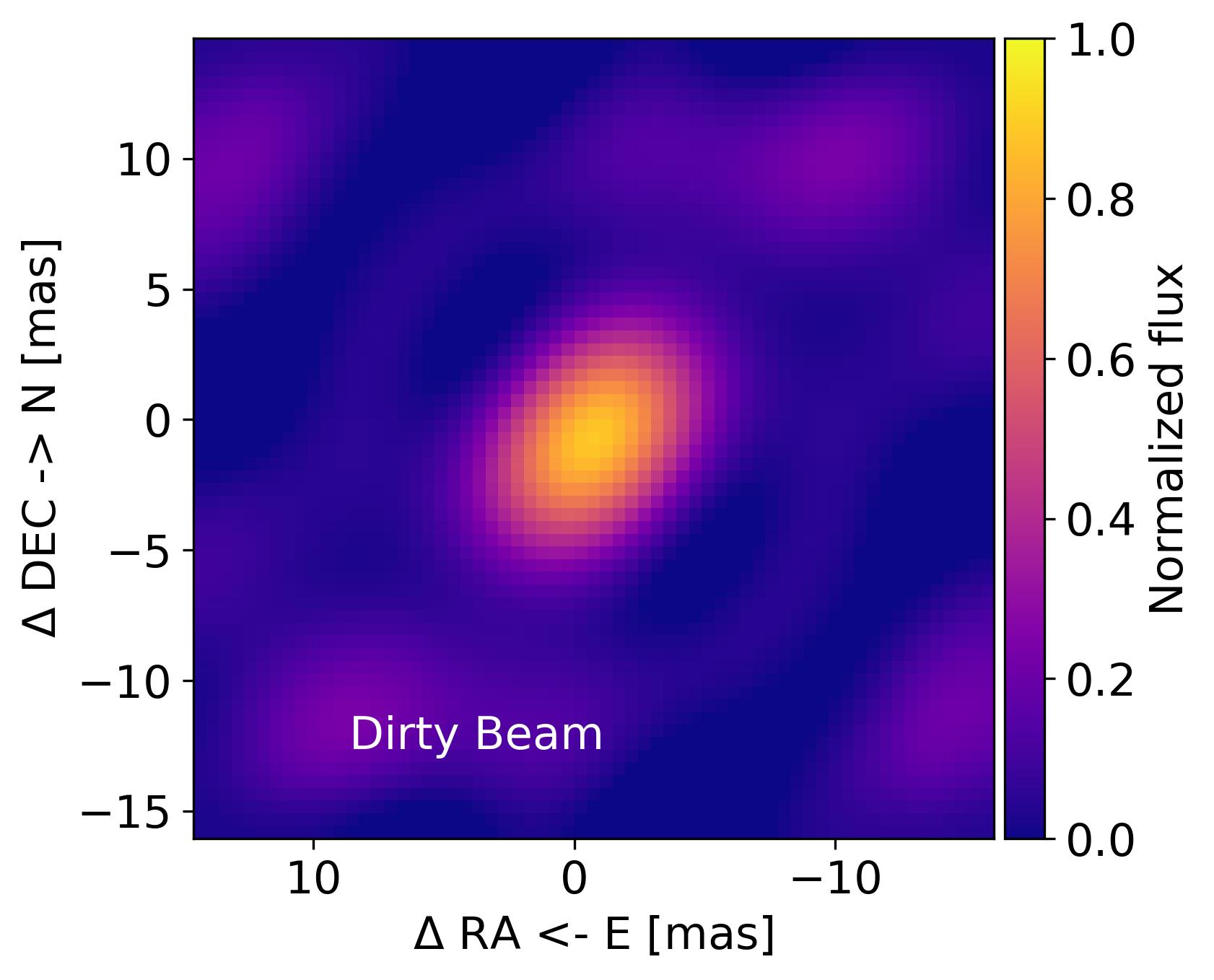} \\
  
   \end{tabular}
   \end{center}
   \caption{ Image reconstruction of RT Pav from PIONIER, GRAVITY, and MATISSE covering continuum wavelengths between 1.5-5.0$\mu$m with a pixel resolution set at $\lambda$/4B$_{max}$. The bottom row shows the dirty beams.}

   { \label{fig:image_reco_all} }
 \end{figure*}

\begin{figure} [h!]
   \begin{center}
    \includegraphics[height=5.2cm]{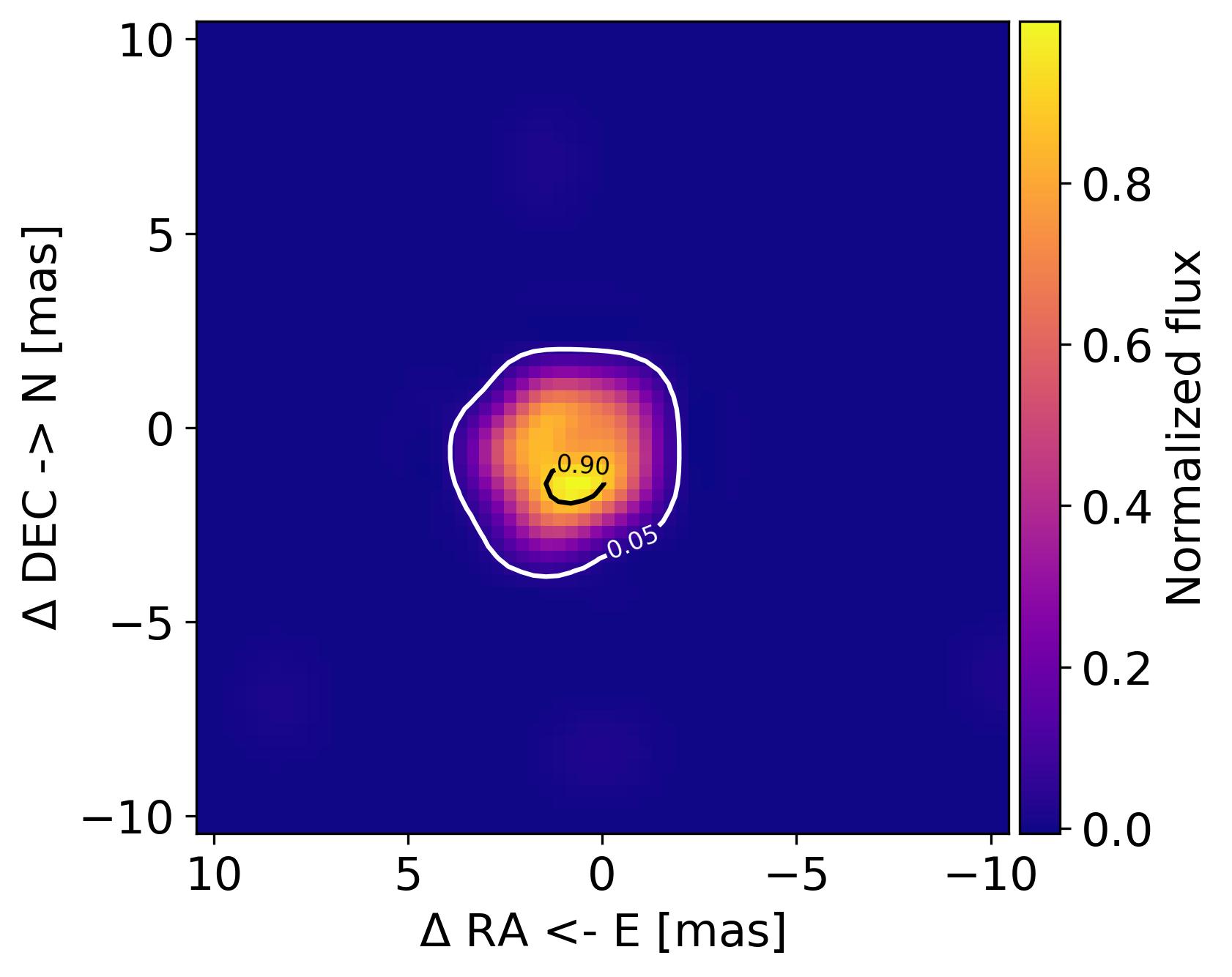} 
   \end{center}
   \caption{ Narrow band near infrared image reconstructions of RT Pav observed at the CO4-2 band head with GRAVITY using the \texttt{MiRA} software.  } 
   \label{fig:CO_bandhead_img}
 \end{figure}

Aperture synthesis was done primarily using the Multi-aperture Image Reconstruction Algorithm (\texttt{MiRA}). While the results of the \texttt{MiRA} reconstructions are reported here, other algorithms and images were also explored and compared using the \texttt{OImaging} software \citep{oimaging_2022SPIE}. The essence of the \texttt{MiRA} reconstruction algorithm is to minimize a cost function \(J(I) = \chi^2(I) + \mu R(I)\), where \(\chi^2(I)\) measures how well the reconstructed image ($I$) fits the interferometric data (e.g., visibilities, closure phases), \(R(I)\) imposes prior constraints such as smoothness, sparsity, or edge-preserving features depending on the specific regularization used, and $\mu$ is the user defined regularisation weight. \texttt{MiRA} employs an iterative approach using gradient-based optimization techniques to minimize the cost function, ensuring convergence toward an image that both fits the data and satisfies the regularization to avoid fitting noise. Specific details of its implementation can be found in  \citet{MIRA_Thiebaut2008, MIRA_Renard2011}. \\\\
The pixel scale is set to \(\lambda / 4B_\text{max}\), where \(\lambda\) is the central wavelength and \(B_\text{max}\) is the maximum baseline of the interferometric array, ensuring appropriate sampling of the spatial frequency domain as dictated by the array's resolution limit. The GRAVITY images were reconstructed in a continuum waveband between 2.1-2.2$\mu m$. 
Priors for the images were created from the best-fit parametric models presented in Tables \ref{tab:baseline_parameter_results}. Final images displayed have been over sampled and smoothed with a Gaussian near the dirty beams FWHM. To better understand the noise properties of the reconstruction algorithm and how the hyper-parameters influenced it, a general pipeline was set up where the images from the best-fit parametric models were used to recreate noisy synthetic VLTI data from the RT Pav observations. This was then passed to the image reconstruction algorithm over a parameter grid search to investigate which parameters best reproduced the parametric model from the observations $(u,v)$ coverage. Table \ref{tab:image_reco_parameters} outlines the hyper-parameters used in the final image reconstructions presented in Figures \ref{fig:image_reco_all} and \ref{fig:CO_bandhead_img}.

\begin{table}[h!]
    \centering 
    \caption{Input parameters to \texttt{MiRA} image reconstruction algorithm.}
    \begin{tabular}{|c|c|c|c|c|}
         \hline
         Instrument &  Prior & Reg. & Parameters  \\ \hline
         PIONIER (H) & Binary & Hyperbolic & $\mu$=100, $\tau$=$10^{-5}$ \\ \hline
         GRAVITY (K)  & Ellipse & Hyperbolic & $\mu$=100, $\tau$=$10^{-1}$\\ \hline
         MATISSE (L) & Binary & Hyperbolic & $\mu$=3000, $\tau$=1\\ \hline
         MATISSE (M) & Binary & Hyperbolic & $\mu$=3000, $\tau$=1\\ \hline
    \end{tabular}
    \label{tab:image_reco_parameters}
\end{table}

\section{Discussion}
The combined photometric, astrometric, and interferometric results place tight constraints on any binary interpretation of RT Pav's LSP. As shown in Sect.~3, the Gaia DR3 constraints limit any companion to $M_2 \lesssim 0.1\,M_\odot$, implying a Roche-lobe radius of $\sim0.35$~AU, which is substantially smaller than the projected radius ($r_d \sim 0.7$~AU) required to reproduce the observed light-curve depth. Thus, any obscuring structure large enough to account for the photometric modulation would have to extend beyond the gravitationally bound volume of such a companion.
\\\\
Formally, binary parameterization  often yield the lowest $\chi^2_\nu$ among the simple models; however, several independent lines of evidence argue against a genuine bound companion causing the LSP. First, the inferred companion positions are not coherent across wavelength: separations and position angles differ between bands and instruments, rather than tracing a single astrometric location compatible with the LSP phase and Gaia constraints. Second, Appendix \ref{ap:dust_properties} shows that, under current understanding of O-rich AGB dust properties \citep{hofner_2018_mass_loss_review}, a $\sim$1\,\% near-IR companion flux is in principle achievable (via scattering and/or thermal emission) at LSP-consistent separations, within the Gaia constrained Roche-lobe size, and is slightly above our typically measured $3\sigma$ detection limits. If such a companion existed, one would therefore expect a coherent $\gtrsim3\sigma$ detection across bands. Instead, significant detections ($>3\sigma$, relative to a UD reference) occur only in the PIONIER H band and the MATISSE M band, while GRAVITY K and MATISSE L bands do not show significant companions at comparable contrasts. Third, the pattern of minima in binary fit chi-squared maps correlates with anisotropic $(u,v)$ coverage. Synthetic observations that adopt the real $(u,v)$ sampling and noise but contain only photospheric structure (UD or oscillatory dipole) reproduce similar, centro-symmetric $\chi^2$ trenches in the 3–5\,mas range (Appendix \ref{ap:binary_grid_search}), indicating that the PIONIER-scale detections can arise from sampling-driven aliasing rather than genuine off-axis flux. Consistently, when the best-fit oscillatory convective dipole model (from the H band) is divided out of the data and the residuals are refitted, no companion is recovered above the 3$\sigma$ flux detection limits (non-detection), further disfavoring a real secondary as the cause of the LSP. It is important to note, however, that this does not exclude the existence of a companion; rather, it suggests that any such object is unlikely to be the primary driver of RT Pav’s LSP.
\\\\
Where the photosphere is resolved (PIONIER H band), an oscillatory convective dipole with a temperature contrast of $\sim$200\,K reproduces both the interferometric morphology and, independently, the amplitude and waveform of the visible LSP light curve. Also the observed decrease of closure-phase amplitude with wavelength is consistent with a declining temperature contrast at longer wavelengths as shown in Fig. \ref{fig:mc_thermal_v_binary}. Image reconstructions across all bands show consistent photospheric asymmetries without repeatable off-axis emission at similar contrast levels, supporting an intrinsic origin of the observed variability rather than a companion-driven scenario. It was also noticed during the fitting process that dipole modes oscillating near perpendicular to the line of sight can produce the symmetrical secondary maxima in the light curves which is an observable test for the oscillatory convective dipole mode which contrasts to the secondary minima discussed by \citet{soszyski_2021_binary_origin_LSP} to be evidence that binarity is the cause of the LSP.  An example of this is shown in Appendix \ref{ap:secondary_maxima_dipole}. Future studies will investigate the full parameter space of this phenomenon.  While the oscillatory convective dipole model provides a physically self-consistent explanation that does not violate the Gaia astrometric or photometric constraints, it could only be reliably fitted to the PIONIER H band dataset, where the photosphere is resolved, whereas the other interferometric datasets (GRAVITY and MATISSE) lacked sufficient spatial resolution to test the model under comparable conditions. This asymmetry in model applicability introduces a bias in the comparison between the binary and convective-mode hypotheses and underscores the need for multi-epoch, multiband observations at consistent resolution to confirm the temporal and spectral evolution of the inferred dipole asymmetry. 
\\\\
Taken together, violations of geometric and astrometric bounds, lack of coherent astrometric positions, the absence of the expected wavelength trend for dusty companions, and the removal of apparent signals after subtracting the best-fit dipole, the data favor a purely photospheric interpretation of RT Pav’s LSP. Future multi-epoch interferometric with improved $(u,v)$ coverage, and spectro-interferometric monitoring across the LSP cycle will be essential to test temporal coherence of the dipole signature and to quantify any evolution in orientation or amplitude.

\section{Conclusions}
The combined photometric, astrometric, and interferometric constraints place stringent limits on a binary interpretation for RT Pav’s LSP. Although parametric binary fits yield formally low $\chi^2$ values—some with apparent high significance—their fitted positions significantly vary between instruments and wavelengths, often violate the Gaia astrometric constraints for the LSP, and synthetic tests show that photospheric asymmetries can alias into similar companion-like signals under anisotropic $(u,v)$ coverage. Furthermore, the L and M band closure phases remain consistent with zero, contrary to expectations for the dusty companion hypothesis, especially given the observed photometric variability of the LSP. In contrast, an oscillatory convective dipole model with $\sim$200\,K temperature contrast reproduces both the observed H band morphology and independently, the amplitude and waveform of the visible LSP light curve, providing a self-consistent explanation without violating the Gaia or photometric constraints. Additionally, where the stellar surface is resolved (H band), subtracting this best-fit photospheric dipole model eliminates a companion signal above the detection limits. Taken together, the evidence favors an intrinsic, photospheric origin for RT Pav’s LSP. Confirming this interpretation will require multi-epoch, multiband interferometric and spectro-interferometric monitoring to track the temporal evolution of the observed asymmetries.

\begin{acknowledgements}
Part of this work was supported through an Australian Government Research Training Program Scholarship. IEM acknowledges support from the grant ANID FONDECYT 11240206. D. Defr\`ere has received funding from the European Research Council (ERC) under the European Union's Horizon 2020 research and innovation program (grant agreement CoG - 866070). This research has benefited from the help of SUV, the VLTI user support service of the Jean-Marie Mariotti Center (https://www.jmmc.fr/suv). A special thanks to Alexis Matter for the help. Finally, thanks to Samuel Courtney-Barrer for the constant guidance. 
\end{acknowledgements}

\bibliographystyle{aa} 
\bibliography{bib.bib} 

\begin{appendix} 

\section{VLTI observations} \label{ap:vlti_obs}
Tables \ref{tab:pionier_obs} - \ref{tab:mat_obs} outline the observations used for this study under the ESO programs 109.23K5.001, 109.23K5.004, 109.23K5.0067, 109.23K5.008, 109.23K5.009. Table \ref{tab:rt_pav_calibrators} outline the calibrators used for the observations taken on each instrument.
\begin{table}[h]\label{tab:rt_pav_calibrators}
\centering
\caption{Calibrators}
\begin{tabular}{cccc}
\hline\hline
Calibrator Name & Instrument & Mag. & Diam. [mas]\\
HD 171042 & PIONIER / H & 3.6 (H) & 0.87 $\pm$ 0.01\\
HD 181019 & PIONIER / H & 3.4 (H) & 0.99 $\pm$ 0.01\\
HD 172211 & PIONIER / H & 3.5 (H) & 0.87 $\pm$ 0.01\\
HD 181019 & GRAVITY / K & 3.3 (K) & 0.99 $\pm$ 0.01 \\
Zeta Pav & MATISSE / L/M & -1.2 (L) & 2.70 $\pm$ 0.12 \\ 
\hline
\hline
\end{tabular}
\tablefoot{Calibrators used for different instruments / wavebands for the RT Pav observations on the VLTI. Diameter uncertainties are calculated from the standard deviation of GAIA, MIDI, and COHEN measured diameters in MDFC Version 10 \citet{cruzalebes_2019_midIR_calibrators} catalog. If only GAIA diameters were available the uncertainty is assumed to be 1\% of the measured diameter in line with \citet{heiter_2015_gaia_diam}.}
\end{table}

\begin{table}[h!]\label{tab:PIONIER_obs}
    \centering
    \caption{PIONIER observations}
    \begin{tabular}{c||c|c|}
        Instrument & Date & Configuration \\
        PIONIER & 28/4/22T09:22 & Small  \\
        PIONIER & 30/4/22T09:40 & Small \\
        PIONIER & 3/5/22T07:11 & Medium    \\
        PIONIER & 3/5/22T07:47 & Medium    \\
        PIONIER & 3/5/22T08:16 & Medium    \\
        PIONIER & 3/5/22T09:03 & Medium    \\
        PIONIER & 9/6/22T10:11 & Large   \\
        PIONIER & 23/6/22T08:37 & Small  \\
        PIONIER & 23/6/22T09:11 & Small  \\
        PIONIER & 4/7/22T01:16 & Large \\
        PIONIER & 4/7/22T02:39 & Large \\
        PIONIER & 4/7/22T04:09 & Large \\
        PIONIER & 4/7/22T04:48 & Large\\
        PIONIER & 20/8/22T04:15 & Small  \\
        PIONIER & 22/8/22T01:42 & Small 
    \end{tabular}
    \tablefoot{8.4hrs PIONIER. date day corresponds to beginning of night, while the timestamp is rounded down to the  minute (UT time) at beginning of 1st science template in CAL-SCI-CAL-SCI-CAL concatenation.  }
    \label{tab:pionier_obs}
\end{table}

\begin{table}[h!]\label{tab:GRAVITY_obs}
    \centering
    \caption{GRAVITY observations}
    \begin{tabular}{c||c|c|}
        Instrument & Date & Configuration  \\
        GRAVITY & 23/6/22T07:41 & Small \\
        GRAVITY & 26/8/22T02:42 & Medium \\
        GRAVITY & 2/9/22T06:47 & Large \\
        GRAVITY & 5/9/22T06:09 & Astrometric  \\

    \end{tabular}
    \tablefoot{4Hrs GRAVITY observations. }
    \label{tab:gravity_obs}
\end{table}

\begin{table}[h!]\label{tab:MATISSE_obs}
    \centering
    \caption{MATISSE observations}
    \begin{tabular}{c||c|c|}
        Instrument & Date & Configuration  \\
        MATISSE$^*$ & 7/4/22T06:58 & Large \\
        MATISSE$^*$ & 14/5/22T05:34 & Large \\
        MATISSE$^*$ & 14/5/22T06:47 & Large \\
        MATISSE$^*$ & 16/5/22T06:29 & Large  \\
        MATISSE$^*$ & 16/5/22T07:39 & Large  \\
        MATISSE$^{***}$ & 2/6/22T06:20 & Medium \\
        MATISSE & 12/7/22T01:07 & Large \\
        MATISSE$^{***}$ & 27/7/22T23:23 & Small  \\
        MATISSE & 27/7/22T00:46 & Small  \\
        MATISSE & 27/7/22T01:59 & Small \\
        MATISSE & 27/7/22T04:34 & Small \\
        MATISSE & 28/7/22T23:42 & Large \\
        MATISSE$^{**}$  & 30/7/22T01:59 & Medium\\
        MATISSE & 30/7/22T03:08 & Medium  \\
        MATISSE & 1/8/22T23:57 & Medium  \\
        MATISSE & 1/8/22T04:01 & Medium\\
        MATISSE$^{***}$  & 5/8/22T01:05 & Large\\
        MATISSE & 27/8/22T23:26 & Medium  \\
        MATISSE & 30/8/22T23:45 & Large   \\
        MATISSE & 30/8/22T23:45 & Large \\
        MATISSE & 30/8/22T02:02 & Large \\
    \end{tabular}
    \tablefoot{18Hrs MATISSE. Date day corresponds to beginning of night, while the timestamp is rounded down to the  minute (UT time) at beginning of science template, * N-band data is in-valid due to bad N-band calibrator HD 157045. ** - intermediate configuration K0-G2-D0-J3. *** - Excluded from analysis due to data quality. }
    \label{tab:mat_obs}
\end{table}

\section{Analytic ellipse model} \label{ap:ellipse_model}
Considering an analytic expression of an object $f(\boldsymbol{x})$ with coordinates $\boldsymbol{x}= (x,y)$ we may consider a general affine transform of the form 
\begin{equation}
    \boldsymbol{x'} = \boldsymbol{  A  x + x_0}
\end{equation}
defining g as the affine transformed function f: 
\begin{equation}
    g(\boldsymbol{x}) = f(\boldsymbol{  A  x + x_0})
\end{equation}
Using the shift and scaling properties of the Fourier transform $f( \boldsymbol{x}) \xrightarrow[]{F} \hat f(\boldsymbol{u})$ we can analytically find the Fourier transform of g:
\begin{multline} \label{eq:affine_transform_visibility}
    \hat g(u,v) = \frac{1}{|A|} \exp{\left( \frac{i2\pi}{|A|}((dx_0 - by_0)u + (ay_0 - cx_0)v)\right)} \\ \hat f\left( \frac{du-cv}{|A|} , \frac{av-bu}{|A|} \right) 
\end{multline}
where $|A|=det(A)=ad-bc$. In compact notation:
\begin{equation}
    \hat g(u,v) = \frac{1}{|A|} e^{2\pi i \boldsymbol{x_0^T A^{-T} u}} \hat f(\boldsymbol{A^{-T} u})
\end{equation}
where $\boldsymbol{A^{-T}}$ is the inverse transpose of $\boldsymbol{A}$. i.e:
\begin{equation}
    \boldsymbol{A^{-T}} = \frac{1}{|A|}
    \begin{bmatrix}
    d & -c\\
    -b & a
    \end{bmatrix} 
\end{equation}
Any non-singular 2D linear transform $\boldsymbol{A}$ converts a disk to an ellipse. Therefore using the Fourier transform properties of an affine transformation defined in Eq. (\ref{eq:affine_transform_visibility}), we can use the analytic expression for the centered uniform disk visibility (Eq. (\ref{eq:ud_disk_visibility})) to derive the visibility of a centered ellipse that's transformed via $\boldsymbol{A}$:
\begin{equation} \label{eq:ellipse_visibility}
    V(u,v) = \frac{1}{ad-bc} \frac{2J_1\left( \pi \theta \sqrt{\left(\frac{du-cv}{ad-bc}\right)^2 + (\frac{av-bu}{ad-bc})^2 }\right)}{ \pi \theta \sqrt{(\frac{du-cv}{ad-bc})^2 + (\frac{av-bu}{ad-bc})^2 } } 
\end{equation}

\section{H band binary model grid fit }\label{ap:binary_grid_search}
To assess the robustness of binary detections and the potential for false companion signals arising from interferometric sampling, 
we performed a grid search of unresolved companion models in the H band using \texttt{CANDID} and \texttt{PMOIRED}.  Synthetic datasets were generated for purely photospheric geometries (uniform disk and oscillatory convective dipole)  with the same $(u,v)$ coverage and noise properties as the real PIONIER observations.  Comparing these synthetic fits with those to the actual data allows us to evaluate how anisotropic $(u,v)$ sampling and limited phase coverage can mimic significant $\chi^2$ minima in the 3–5\,mas range, where candidate companions were nominally detected.
\begin{figure*}[h]
    \centering
    \includegraphics[width=18cm]{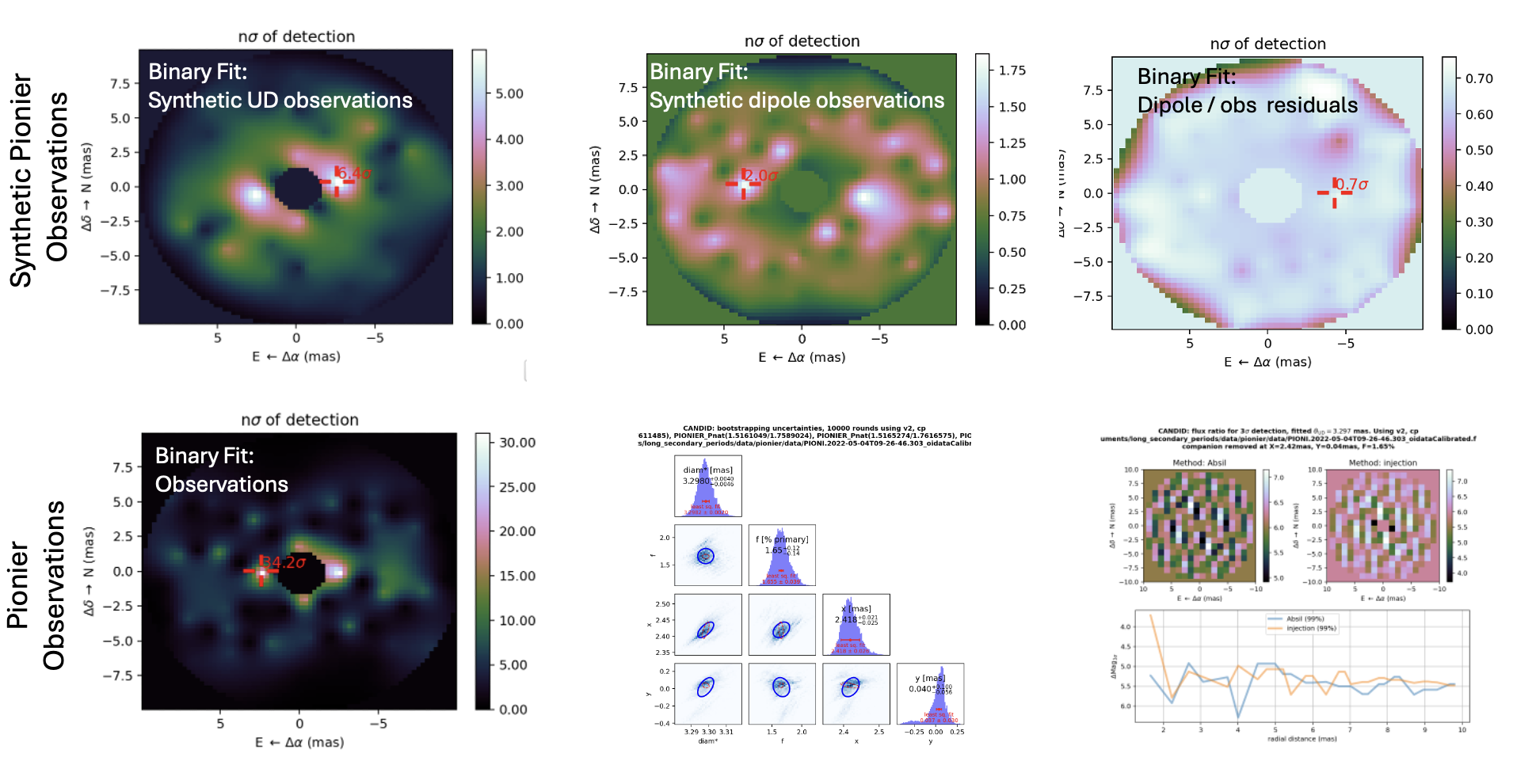}
    \caption{ PIONIER Fit Maps using \texttt{CANDID} \citep{gallenne_2015_candid}. The top row shows fits to synthetic PIONIER observations of UD, oscillatory convective dipole (keeping the respective observed uncertainties), and residuals of the  oscillatory convective dipole model with real observations. The bottom row shows the binary fit to the real data along with parameter distributions from bootstrapping and detection limit estimates. }
    \label{fig:PIONIER_binary_gridsearch}
\end{figure*}

\section{Constraining companion dust properties} \label{ap:dust_properties}
Even for the most massive companion allowed by the Gaia constraint ($M_2\lesssim0.1\,M_\odot$), the intrinsic photospheric contrast is far too small to explain the typical $\sim1\%$ flux ratios fitted when considering the Stefan–Boltzmann law. Hence the fitted flux ratios cannot be intrinsic companion light and must arise from circumsecondary or circumbinary dust through scattering and/or thermal emission.
The near-infrared excess may arise either from scattering of stellar photons or from thermal emission by hot circumsecondary dust. In the single–scattering, optically thin limit the fractional scattered flux is
\begin{equation}
\frac{F_{\mathrm{sca}}}{F_\star} \simeq \frac{C_{\mathrm{sca}}}{4\pi a^2}\,\Phi(\alpha),
\label{eq:Fsca}
\end{equation}
where \(C_{\mathrm{sca}}=\pi s^2 Q_{\mathrm{sca}}\,N_d\) is the total scattering cross-section of \(N_d\) grains of radius \(s\) and efficiency \(Q_{\mathrm{sca}}\); \(a\) is the star–dust separation; and \(\Phi(\alpha)\) is the phase function (normalized to \(\int \Phi\,\mathrm d\Omega=4\pi\)). Writing the dust mass as \(M_d = (4\pi/3)\,\rho_g s^3 N_d\) (material density \(\rho_g\)), we obtain
\begin{equation}
M_d = \frac{4\,\rho_g s}{3\,Q_{\mathrm{sca}}}\,C_{\mathrm{sca}}.
\label{eq:Md_from_Csca}
\end{equation}
Combining Equations~\eqref{eq:Fsca}–\eqref{eq:Md_from_Csca} gives, for a target fraction \(\epsilon\equiv F_{\mathrm{sca}}/F_\star\),
\begin{equation}
C_{\mathrm{sca}} = \frac{\epsilon\,4\pi a^2}{\Phi(\alpha)},
\qquad
M_d = \frac{4\,\rho_g s}{3\,Q_{\mathrm{sca}}}\,\frac{\epsilon\,4\pi a^2}{\Phi(\alpha)}.
\label{eq:sca_limits}
\end{equation}
Adopting \(a=1.6~\mathrm{AU}\) (\(\simeq3.17\) mas at 505 pc), \(\epsilon=0.01\), \(\Phi(\alpha)=1\) (isotropic), \(s=0.2~\mu\mathrm m\), \(\rho_g=3~\mathrm{g\,cm^{-3}}\), and \(Q_{\mathrm{sca}}=1\) yields
\[
C_{\mathrm{sca}}\simeq 0.32~\mathrm{AU}^2,\qquad
M_d \simeq 3\times10^{-12}\,M_\odot,
\]
with \(M_d \propto s\,\rho_g/Q_{\mathrm{sca}}\) and \(M_d \propto 1/\Phi(\alpha)\) (forward scattering lowers the mass linearly by \(\Phi\)). These relations follow classical radiation–pressure and Mie formulations \citep{burns_1979_radiation_force_particles,draine_1984_mie_scat_silicate_grains_properties}, and the grain sizes/locations are consistent with Fe-poor, Mg-rich silicates forming at \(\sim2R_\star\) with \(s\sim0.1\)–\(1~\mu\mathrm m\) \citep{hofner_2008_wind_M-type_agb,hofner_2018_mass_loss_review}.

For thermal emission, the band contrast can be written as
\begin{equation}
\frac{F_{\nu,\mathrm{th}}}{F_{\nu,\star}}
\simeq \frac{C_{\mathrm{abs}}}{\pi R_\star^2}\,
\frac{B_\nu(T_{\rm eq})}{B_\nu(T_\star)},
\qquad
C_{\mathrm{abs}}=\pi s^2 Q_{\mathrm{abs}}\,N_d
= \frac{3\,Q_{\mathrm{abs}} M_d}{4\rho_g s},
\label{eq:Fth}
\end{equation}
where \(Q_{\mathrm{abs}}\) is the absorption efficiency and \(B_\nu\) is the Planck function \citep{draine_1984_mie_scat_silicate_grains_properties}. For (approximately) gray grains in radiative equilibrium,
\begin{equation}
T_{\rm eq}\simeq T_\star\left(\frac{R_\star}{2a}\right)^{1/2},
\label{eq:Teq}
\end{equation}
consistent with AGB wind and binary temperature profiles \citep{hofner_2008_wind_M-type_agb,hofner_2018_mass_loss_review,mellah_2020_wind_cool_evolved_star}. Inserting Equations~\eqref{eq:Fth}–\eqref{eq:Teq} and solving for \(C_{\mathrm{abs}}\) at a target \(\epsilon_\nu\equiv F_{\nu,\mathrm{th}}/F_{\nu,\star}\) gives
\begin{equation}
C_{\mathrm{abs}} = \epsilon_\nu\,\pi R_\star^2 \left[\frac{B_\nu(T_\star)}{B_\nu(T_{\rm eq})}\right],
\qquad
M_d=\frac{4\,\rho_g s}{3\,Q_{\mathrm{abs}}}\,C_{\mathrm{abs}}.
\label{eq:abs_limits}
\end{equation}
For RT~Pav, taking \(T_\star=3000~\mathrm K\), \(R_\star=0.808~\mathrm{AU}\) (from \(\theta_{\rm UD}=3.2~\mathrm{mas}\)), \(a=1.6~\mathrm{AU}\Rightarrow T_{\rm eq}\!\approx\!1500~\mathrm K\), and \(\lambda=1.65~\mu\mathrm m\) (\(H\) band) gives \(B_\nu(1500~\mathrm K)/B_\nu(3000~\mathrm K)\approx 0.052\). For \(\epsilon_\nu=0.01\),
\[
C_{\mathrm{abs}} \simeq 0.01\,\pi R_\star^2 \left[\frac{1}{0.052}\right] \simeq 0.39~\mathrm{AU}^2,
\]
and, with \(s=0.2~\mu\mathrm m\), \(\rho_g=3~\mathrm{g\,cm^{-3}}\),
\[
M_d \simeq 3.5\times10^{-12}\,Q_{\mathrm{abs}}^{-1}\,M_\odot,
\]
i.e., \(M_d\approx 3.5\times10^{-12}\,M_\odot\) for \(Q_{\mathrm{abs}}=1\), or \(3.5\times10^{-11}\)–\(3.5\times10^{-10}\,M_\odot\) for \(Q_{\mathrm{abs}}=0.1\)–\(0.01\). These expressions provide transparent bounds on the cross-sections, equilibrium temperatures, and dust masses required to reach a \(\sim\)1\,\% \(H\)-band contribution at the companion’s separation.
\\\\
The derived dust masses ($M_d\!\sim\!10^{-12}\!-\!10^{-10}\,M_\odot$) are consistent with the amounts that could plausibly accumulate around a low-mass companion embedded in an AGB wind. 
For a typical O-rich AGB star with gas mass-loss rates 
$\dot M_{\mathrm{gas}}\!\sim\!10^{-7}$--$10^{-6}\,M_\odot\,\mathrm{yr^{-1}}$
and a dust-to-gas ratio 
$\psi\!\sim\!10^{-3}$--$10^{-2}$,
the corresponding dust production rate is
$\dot M_{\mathrm{dust}}\!\sim\!10^{-9}$--$10^{-8}\,M_\odot\,\mathrm{yr^{-1}}$.
A substellar companion of $M_2\!\lesssim\!0.1\,M_\odot$ orbiting at $a\simeq1.6$\,AU 
would capture only a small fraction of this outflow through Bondi--Hoyle accretion or wind--Roche-lobe overflow. 
For relative velocities $v_w\!\sim\!10\,\mathrm{km\,s^{-1}}$, 
the capture efficiency is $f_{\mathrm{cap}}\!\sim\!(R_{\mathrm{acc}}/2a)^2\!\sim\!10^{-2}$--$10^{-3}$,
yielding a dust capture rate of 
$\dot M_{\mathrm{dust,cap}}\!\sim\!10^{-12}$--$10^{-11}\,M_\odot\,\mathrm{yr^{-1}}$.
If grains survive for a timescale comparable to the orbital period ($P\simeq2$\,yr), 
the steady-state dust mass retained in the companion's vicinity is then
$M_{d,\mathrm{steady}}\!\sim\!\dot M_{\mathrm{dust,cap}}P\!\sim\!(2$--$6)\!\times\!10^{-12}$--$10^{-11}\,M_\odot$,
matching the range derived above from the scattering and thermal-emission analysis.
\\\\
For the adopted $a=1.6$\,AU and $q=0.1$, the companion Roche-lobe radius is $R_{L,2}\!\simeq\!0.35$\,AU.
The corresponding effective cross-section required to reproduce a $\sim$1\,\% flux ratio
($C_{\mathrm{sca}}\!\approx\!0.3$--$0.4$\,AU$^2$) implies a dust envelope radius comparable to $R_{L,2}$,
consistent with a wind-fed, Roche-lobe-filling configuration. However, the high equilibrium temperature ($T_{\mathrm{eq}}\!\simeq\!1500$\,K) limits grain survival to only a few orbital timescales, implying that continuous replenishment would be required. Thus, while the estimated masses are compatible with transient, wind-fed envelopes, sustained near-infrared emission at the observed level is difficult to reconcile without more detailed modeling of dust production and confinement.

\section{Secondary maxima in the LSP from an inclined dipole}
\label{ap:secondary_maxima_dipole}
 \begin{figure} [h]
    \includegraphics[width=8cm]{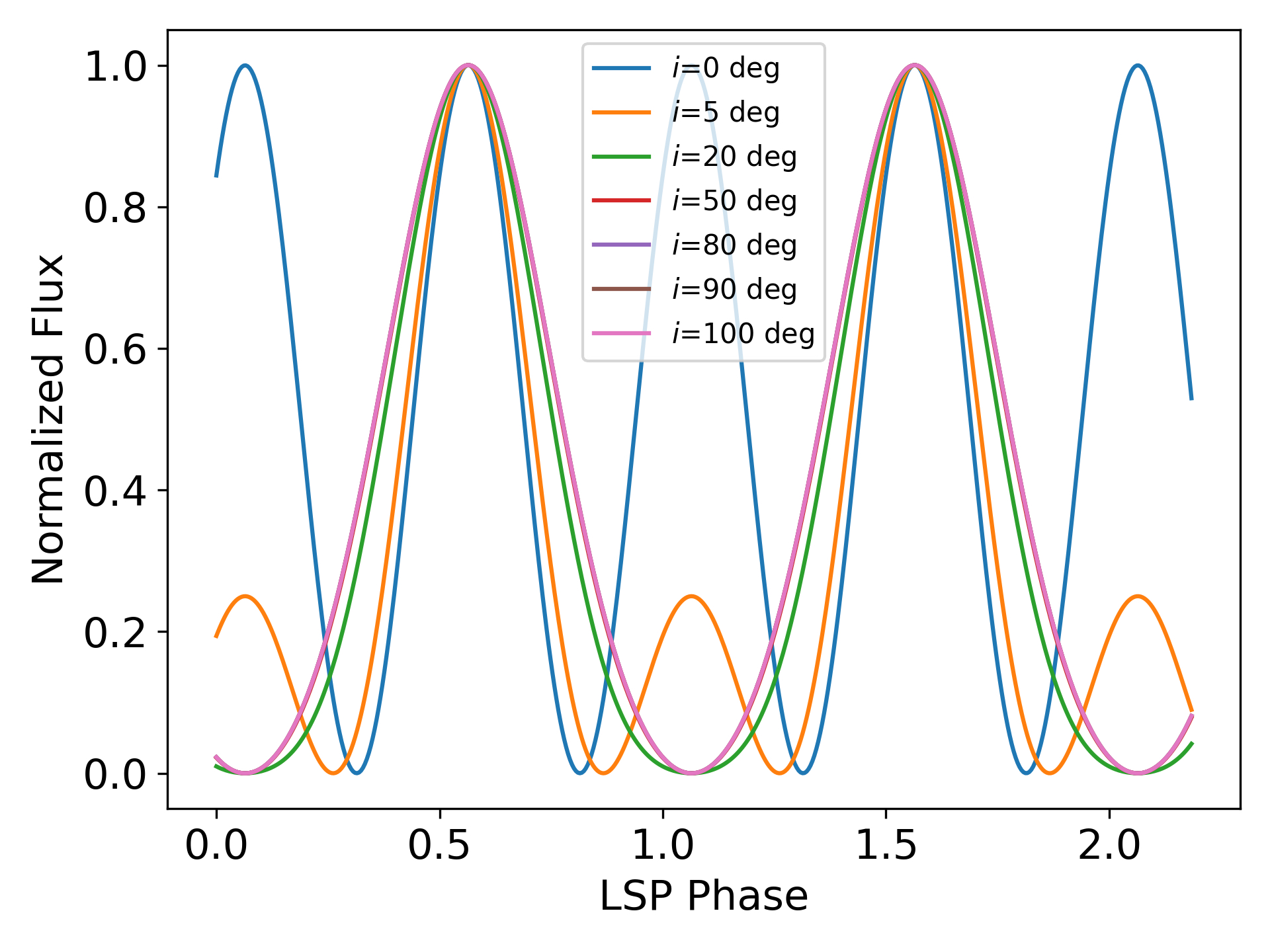}
   \caption{Examples of the oscillatory convective dipole mode causing secondary maxima in the visible light curve when the dipole axis is near perpendicular to the line of sight and the observed wavelength is near the black-body peak from the stellar temperature. Here we considered $T_{eff}$ = 3000K, dipole temperature amplitude = 300K with limb-darkening. } 
   \label{fig:secondary_ecclipse_thermal_dipole}
 \end{figure}
\end{appendix}

\end{document}